\documentclass[a4paper,fleqn,usenatbib]{mnras}
\usepackage[T1]{fontenc}
\usepackage{ae,aecompl}
\usepackage{graphicx}
\usepackage{amsmath}
\usepackage{amssymb}

\title[Planetary nebula NGC 6572]{Morpho-kinematic and photoionization models of the multipolar structures in planetary nebula NGC 6572
}

\author[Bandyopadhyay et al.]{
Rahul Bandyopadhyay$^{1}$,\thanks{E-mail: rahul@bose.res.in}
Ramkrishna Das$^{1}$,
Mudumba Parthasarathy$^{2}$ and
\newauthor{Subhajit Kar$^{1}$} 
\\
$^{1}$S. N. Bose National Centre for Basic Sciences, Block JD, Sector III, Salt Lake, Kolkata 700106, India.\\
$^{2}$Indian Institute of Astrophysics, II Block, Koramangala, Bangalore 560 034, Karnataka, India.\\
}

\date{Accepted XXX. Received YYY; in original form ZZZ}

\pubyear{0000}

\begin{document}
\label{firstpage}
\maketitle

\begin{abstract}

We have studied the planetary nebula (PN) NGC 6572 through 3D morpho-kinematic and photoionization modelling. The 3D morphology is reconstructed from the \textit{Hubble Space Telescope} images in different narrow band filters and position-velocity spectra. The PN have a multipolar morphology consisting of highly collimated outflows. The nebular image show signatures of multiple lobes within a spiral-ring-like structure. The multipolar structure is modelled with two bipolar shells (axes ratio $\sim5.5:1$ and $\sim3:1$), having closed and opened lobes, respectively. A toroidal structure (radius:height $\sim1:3$) surrounds the shells at the waist. The toroidal axis aligns with the major axes of the bipolar shells. Our study reveals the nebula to have a history of collimated polar outflow perpendicular to a higher density equatorial wind with the outflow seemingly have episodes of changing direction of ejection. We construct a photoionization model of NGC 6572 using the deep optical spectra obtained at the 2 m Himalayan Chandra Telescope. For the photoionization model, we configure the input shell geometry in form of a highly bipolar nebular shell with reference to the 3D morphology. Our photoionization model satisfactorily reproduces the observables. We estimate the nebular elemental abundances, and important characteristic parameters of the central star (e.g., effective temperature, luminosity, gravity, mass, etc.) and the nebula (e.g., hydrogen density profiles, radii, etc.). We compare the resolved H$\beta$, [O~{\sc iii}], and [N~{\sc ii}] profiles in the 4.2 m William Herschel Telescope with that from the photoionization model and find a good characteristic match.  

\end{abstract}


\section{Introduction}

Planetary nebulae (PNe) form around low- to intermediate-mass ($\sim1-8\:M_{\sun}$) stars during their late evolutionary phases through interacting stellar winds \citep{1978ApJ...219L.125K}. A large number of PNe exhibit a variety of non-spherical morphologies, with one or more pairs of lobes, forming bipolar, quadrupolar or multipolar morphologies. The morphologies might form due to the directional mass ejection during later phases of stellar evolution and also from the varying dynamics in the outflows afterwards (e.g., \citealt{1989IAUS..131...83B} and references therein). Further, it has been established through studies that the central star binarity might play an important role in shaping of nebular morphologies among PNe, through common envelop evolution (e.g, \citealt{2018ApJ...860...19G}) and formation of precessing outflows (e.g., \citealt{2021ApJ...909...44G}).

The physical quantities associated to asymmetric PNe are subjected to vary along the azimuthal direction due to changing shell configuration. Hence, it is difficult to analyse an asymmetric PN using simplified assumptions, such as spherical shells. For example, a spherical photoionization model of a highly bipolar PNe may not properly reproduce the ionization states, since the ionization of an element depends on the distance from the ionizing source as well as density of the nebular region. Photoionization models of individual PN considering the 3D geometrical characteristics have been computed earlier. For example, IC418 \citep{2009A&A...507.1517M}, NGC 6153 \citep{2011MNRAS.411.1035Y}, Abell 14 \citep{2016MNRAS.457.3409A}, and Tc 1 \citep{2019MNRAS.490.2475A}.  
 
NGC 6572 has shown complex morphological features through various observations. \citet{1999ApJ...520..714M} suggested that the nebula has an elongated structure with central toroid, with signatures of an interacting pair of bipolar shells and a toroidal waist. \citet{2001ApJ...547..302G} reported signatures of two pairs of knots. \citet{2016MNRAS.462..610R} reports concentric rings and arcs around the central nebula. Distance to NGC 6572 has been estimated as $1.18{\pm}0.34$ kpc by \citet{1995AJ....109.2600H}, 1.681 kpc  by \citet{2010ApJ...714.1096S}, $1.46{\pm}0.42$ kpc by \citet{2016MNRAS.455.1459F}, $1.767^{+0.149}_{-0.147}$ kpc by \citet{2021AJ....161..147B}. NGC 6572 has shown signatures of variability (e.g., \citealt{2014ARep...58..702A} and references therein). \citet{1968IAUS...34..376K} first suggested the variability of the PN as evident from the increase in the temperature of the central star. After 20 years, \citet{1988A&A...207L...5M} reported the increase of He~{\sc ii} 4686 {\AA} flux and suggested further increase of central star temperature. 

In this work, we have studied the PN NGC 6572 ($\mathrm{R.A.}=18^h12^m06.31^s$, $\mathrm{Dec.}=+06^{\circ}51^{\prime}13^{\prime\prime}.03$, in epoch J2000) through 3D morpho-kinematic and photoionization modelling analyses. To obtain more detailed knowledge of the shells, we need to obtain a 3D morpho-kinematic structure of the collimated outflows. Furthermore, in these collimated ionized shells, the relative strengths of the ionization states presumably have a $\theta$ variation. Hence, to obtain a proper ionization structure, we may need to treat the collimated morphology of the shells in photoionization models. We construct a 3D morpho-kinematic model of the collimated nebular outflows in NGC 6572 using the 3D modelling code \textsc{shape} \citep{2011ITVCG..17..454S}. Then, we obtain a photoionization model of the PN using the photoionization code \textsc{cloudy} \citep{1998PASP..110..761F,2017RMxAA..53..385F}. We use the code library \textsc{pycloudy} \citep{2013ascl.soft04020M} to include the effects of non-spherical nebular geometry in our photoionization model. Although a comprehensive photoionization model includes dust and molecules (e.g., \citealt{2017ApJS..231...22O,2018A&A...617A..85G,2021MNRAS.503.1543T,10.1093/mnras/stab860,2022MNRAS.509..974G}), we do not include them in our models in this work. We only use optical images and spectra for this work and focus on the structure of the ionized region, as our primary aim is to incorporate the bipolarity of the nebular structure in the nebula of NGC 6572, and hence, derive a morphologically consistent photoionization model of the PN.

In Section 2, we describe the observational data set used for this work. We present the detailed results from our analyses, and the robustness and caveats of our results in Section 3. Finally, we summarise and conclude our paper in Section 4.        

\begin{table}
\centering
\small
\caption{Log of spectroscopic observations. \label{tab:logobs}}
\begin{tabular}{c c c c}
\multicolumn{4}{l}{HCT observations}\\
\hline
Observation date & Slit P.A. ($^{\circ}$) & Grism & Exposure (s)\\
\hline
{ 2021 May 29} & { 0} & { Gr. 7} & ($300\times3$) and 5\\
& & { Gr. 8} & ($300\times3$) and 5\\
& { 90} & { Gr. 7} & ($300\times3$) and 5\\
& & { Gr. 8} & ($300\times3$) and 5\\
{ 2022 June 25} & { 0} & { Gr. 7} & { 20}\\
& & { Gr. 8} & { 20}\\ 
& { 90} & { Gr. 7} & { 20}\\
& & { Gr. 8} & { 20}\\
\hline
&&&\\
\multicolumn{4}{l}{WHT observations}\\
\hline
Observation date & Slit P.A. ($^{\circ}$) & Range ({\AA}) & Exposure (s)\\
\hline
2002 Sep. 13 & 0 & $4290-4735$ & 300\\
& & $4760-5553$ & 300\\
& & $5607-9015$ & 300\\
\hline
&&&\\
\end{tabular}
\begin{tabular}{c c c}
\multicolumn{3}{l}{SPM observations}\\
\hline
Observation date & Slit P.A. ($^{\circ}$) & Exposure (s)\\
\hline
2006 Jul. 13 & 342 & 180\\
2003 Aug. 16 & 92 & 1800\\
\hline
\end{tabular}
\end{table}

\begin{figure}
\centering
\scalebox{1.8}[1.8]{\includegraphics{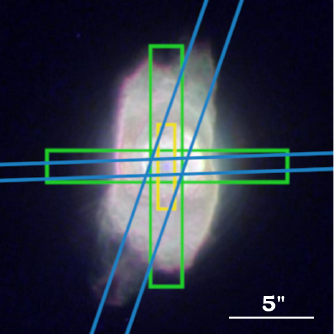}}
 \caption{Colour composite \textit{HST} image of NGC 6572 in H$\beta$ (coloured blue), [O~{\sc iii}] (coloured green), and H$\alpha$ (coloured red), taken through F487N, F502N, and F656N filters, respectively. Spectral extraction windows corresponding to HCT and WHT observations are shown in green and yellow rectangles, respectively. Slits used for obtaining the PV diagrams used in this work are shown using blue lines. \label{fig:ngc6572}}
\end{figure} 

\section{Observational data set} \label{sec:obs}

\subsection{2 m Himalayan Chandra Telescope (HCT) optical spectra} \label{sec:hctspec}

Long-slit spectra of NGC 6572 were obtained using the Hanle Faint Object Spectrograph Camera (HFOSC) at the 2 m Himalayan Chandra Telescope (HCT), Hanle, India. The spectra were obtained in two spectral regions using the grisms Gr. 7, covering $\sim$3700-7000 {\AA} (blue spectral range) with resolution, $R\sim1400$ and Gr. 8, covering $\sim$5500-9100 {\AA} (red spectral range) with $R\sim2200$. The slit-width was $1^{\prime\prime}.92$ and was aligned over the central star of the PN during the observations. We see from the \textit{Hubble Space Telescope (HST)} image that the nebular shells are highly directional and elongated along North-South (NS). Hence, we took the spectra in two slit orientations: East-West (EW) and NS (will be referred as EW spectra and NS spectra, respectively, in the rest of this paper). The observations would mostly cover the waist area and the polar areas, and would map the important areas of radiation. For each orientation, we took longer (300 s) exposure frames to get the weaker lines with good Signal-to-noise ratio. Since the stronger emission lines were saturated by the long exposures, we took short (5 and 20 s) exposure frames to get the stronger emission lines unsaturated. The log of observations is given in Table \ref{tab:logobs}. For flux calibration, spectrophotometric standard star Feige 110 and Feige 66 were observed at the nights of observations using a wide slit configuration.

We reduced the spectra using IRAF package \citep{1993ASPC...52..173T} following the standard reduction techniques as follows. The spectra were bias subtracted and cosmic-ray removed. We extracted the 1D spectra from the 2D spectral image through aperture selection, background subtraction and tracing. The spectra were calibrated in wavelengths using the reference lamp spectra. We calibrated the spectra in absolute flux units using the standard star data. 
 
\subsection{\textit{Hubble Space Telescope (HST)} Images}

We obtained the high-resolution \textit{HST} colour composite image data of NGC 6572 from the Hubble Legacy Archive (HLA\footnote{\url{https://hla.stsci.edu/}}) (PI: Garnett; Proposal ID: 9839; Obs. date: { 2003 September 3}). We further processed the image data of different contrast levels available within HLA to improve the visibility of the fainter structures within the nebula (Fig. \ref{fig:ngc6572}). The composite image maps the narrow band emission from H$\beta$, [O~{\sc iii}], and H$\alpha$, taken through the filters F487N (${\lambda}_p=4865{\AA}$, $\Delta\lambda=26{\AA}$), F502N (${\lambda}_p=5010{\AA}$, $\Delta\lambda=65{\AA}$), and F656N (${\lambda}_c=6564{\AA}$, $\Delta\lambda=22{\AA}$), respectively, using PC chip of the Wide-Field Planetary Camera 2 (WFPC2) instrument. The total exposures corresponding to the H$\beta$, [O~{\sc iii}], and H$\alpha$ images were 360, 240, and 180 s, respectively. The image has a spatial resolution of $0^{\prime\prime}.045$ pixel$^{-1}$.  

\subsection{Position-velocity diagrams}

Position-velocity (PV) diagrams (spatially resolved 2D spectra) are taken from San Pedro M\'artir (SPM) Kinematic Catalogue of Galactic Planetary Nebulae \citep{SPMCatalog}. Long-slit echelle spectroscopic data included in this catalogue were taken at the San Pedro M\'artir National Observatory, Mexico using the Manchester Echelle Spectrometer (MES). For our work, we use the spectra at slit position angles: { $342^{\circ}$} (taken with 2$^{\prime\prime}$ slit, resolution of 11.5 km s$^{-1}$, and 180 s exposure) and { $92^{\circ}$} (taken with 0.$^{\prime\prime}$9 slit, resolution of 9.2 km s$^{-1}$, and 1800 s exposure). The observations were taken on 2006 Jul. 13 and 2003 Aug. 16, respectively. Average seeing was around $1.5^{\prime\prime}$. 

\subsection{4.2 m William Herschel Telescope (WHT) optical echelle spectra}

We use archival optical high-resolution spectra of NGC 6572. The spectra were obtained on 2002 September 13, using the Utrecht Echelle Spectrograph (UES) installed at the 4.2 m William Herschel Telescope (WHT), Roque de los Muchachos Observatory, La Palma, Spain. Spectra were taken in three regions: $4290-4735$, $4760-5553$ and $5607-9015$ {\AA}. Each region was observed with an exposure time of 300 s through a slit of 1 arcsec width. The echelle grating E31 (31.6 lines mm$^{-1}$) and SITe1 CCD ($2048\times2048$ pixels of 24 $\mu$m) were used. The spectrograph provided a resolution of 0.053 {\AA} pixel$^{-1}$. The data were reduced with standard echelle spectroscopy routine in the IRAF package, following the steps of bias and scattered light subtraction, flat-field correction, order extraction and wavelength calibration (using a Th-Ar lamp spectra). The combined wavelength calibrated spectrum ($4300-9000$ {\AA}) was continuum normalized for the present study. \\

In this work, the \textit{\textit{HST}} images and the PV diagrams are used for the 3D morpho-kinematic modelling. HCT optical spectra from our observations are used for the estimation of interstellar extinction, electron temperatures and densities, ionic and total elemental abundances, and to match the flux ratios and absolute flux in the photoionization modelling. WHT spectra is used to match the line profiles from photoionization modelling.

\section{Results and Discussion}

\begin{figure*}
\centering
\scalebox{0.4}[0.4]{\includegraphics{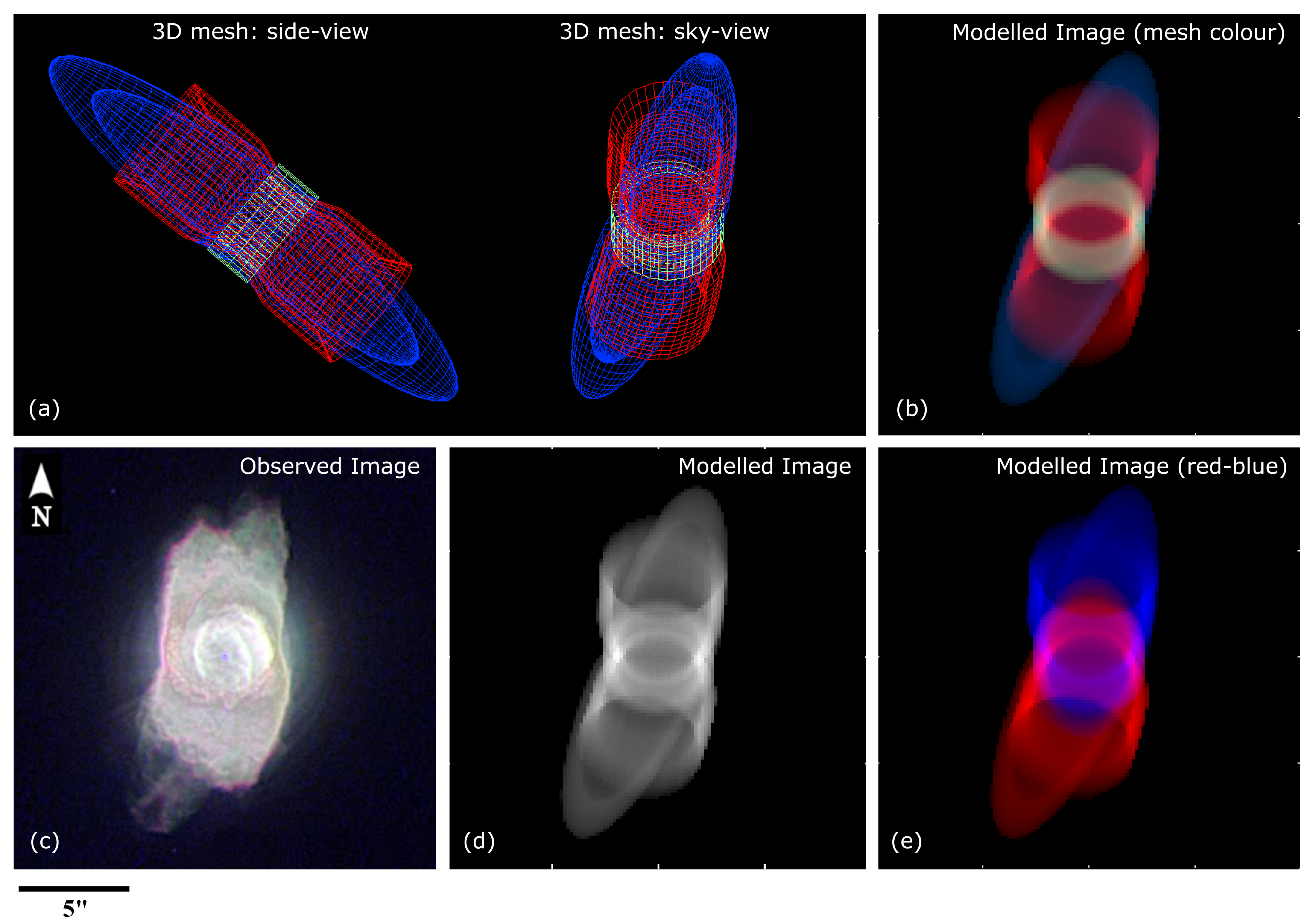}}
 \caption{Reconstructed 3D morphology of NGC 6572 in \textsc{shape}. (a) The 3D mesh consisting of three components an open and a closed lobed bipolar shells, and a torus around the waist of the bipolar shells. The mesh is shown in side-view and sky-view (as seen from the Earth). (b) The coloured 2D modelled image, where the colours correspond to the colour of the mesh, shows the components separately. (c) The observed colour composite \textit{HST} image mapped in emissions of H$\beta$, [O~{\sc iii}], [N~{\sc ii}], and H$\alpha$. (d) Image of the modelled nebula in greyscale, projected in sky-plane, as viewed from the Earth. (e) Modelled red-blue image is shown to depict the red-ward and blue-ward moving regions of the nebula. \label{fig:3dmodel}}
\end{figure*} 

\begin{figure*}
\centering
\scalebox{0.37}[0.37]{\includegraphics{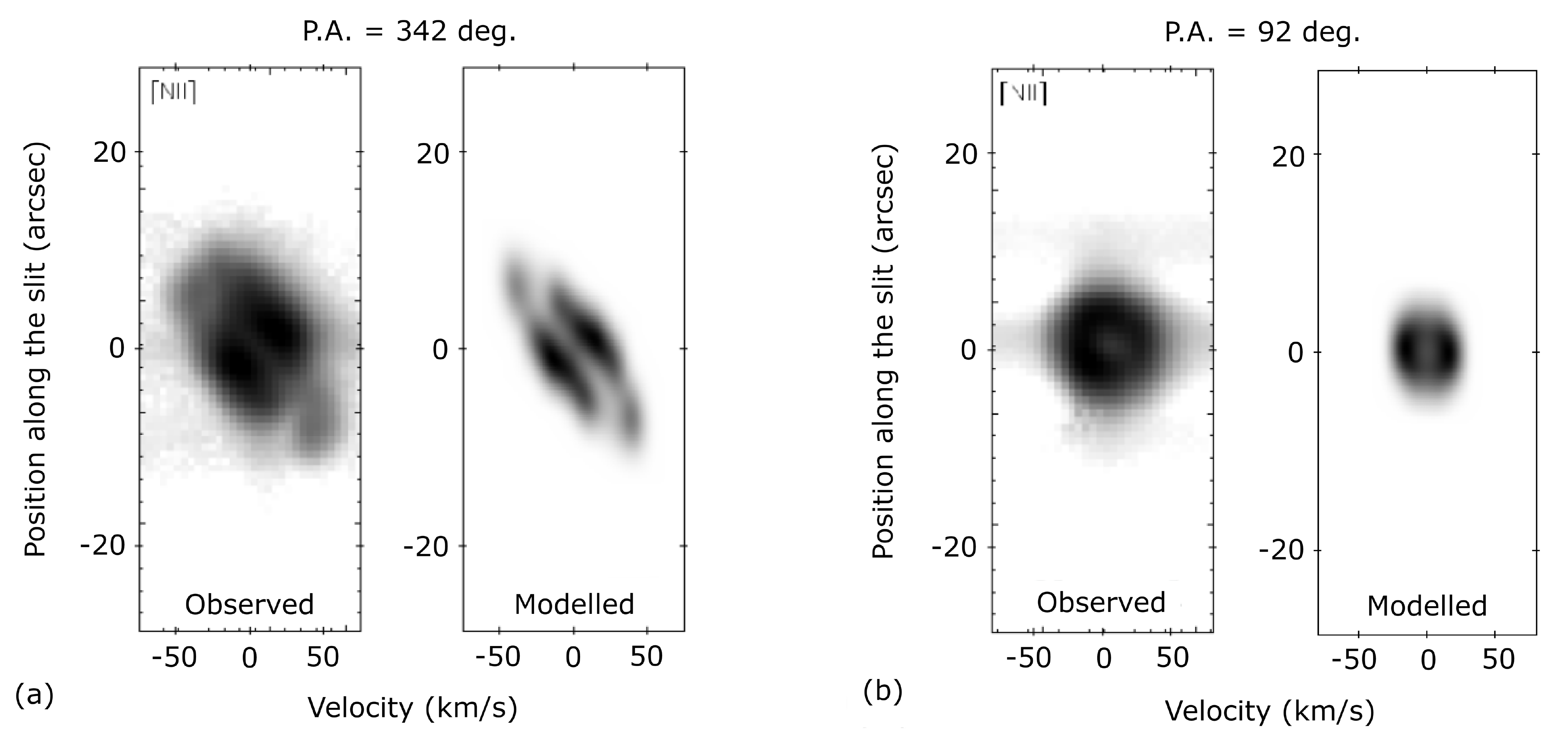}}
 \caption{Comparison between the observed and modelled PV diagrams for position angles $342^{\circ}$ (a) and $92^{\circ}$ (b).  \label{fig:pvdiagrams}}
\end{figure*} 

\subsection{3D morpho-kinematic modelling} \label{sec:3dmkmodel}

\subsubsection{Morphological features from image}

The entire nebular structure has dimensions of $8{\arcsec}{\times}7{\arcsec}$, including the bright and collimated inner shells and the surrounding rings and arcs \citep{2016MNRAS.462..610R}. \citet{1999ApJ...520..714M} concluded about the presence of interacting shells including and a toroidal waist by studying the position-velocity (PV) correlation along the radial direction of the nebula for different position angles. The high resolution \textit{HST} image clearly shows the presence of multiple lobes with different orientations. By further inspection of the \textit{HST} image, we can identify at least two closed lobes in the northern (upper) region. The souther (lower) region shows a fainter closed lobe and a brighter open lobe. However, from the 2D image, we cannot properly identify the connection between the lobes and also their orientations are not clear. 
Through our 3D morpho-kinematic modelling, we attempt to construct the simplest physical structure that might be able to reproduce the main features as described above. We aim to obtain the probable 3D structure of the bright inner shells of the nebula, which apprently shows the presence of multiple lobes in the \textit{HST} image. We do not attempt to model the much fainter rings and arcs, and focus on modelling the collimated outflows, which might give clues to understand the orientation and dynamical evolution of the nebular shells. 

\subsubsection{3D reconstruction of the nebula}

To obtain the full 3D view of the morphology of NGC 6572 and constrain it with the nebular velocity field, we construct a 3D morpho-kinematic model of NGC 6572 using \textsc{shape} \citep{2011ITVCG..17..454S}. Using the interactive interface of \textsc{shape}, we construct the basic 3D mesh of the components of nebular shell using the available basic geometrical structures, e.g., sphere, torus, cone etc. The components are attributed with a `density' parameter that is used to imitate the effects of the physical density of the nebula. Further, a velocity field parameter is used to include the effects of nebular expansion velocity. The code computes radiation transfer through the nebula and generate a synthetic model image (2D projection of the reconstructed model) and PV diagrams that are compared with the observed \textit{HST} image and and the PV diagrams, respectively. We modify the basic geometry in a required way using various structural modifiers, e.g., squeeze, spiral, twist, etc. 
 
To reconstruct the 3D structure with reference to the 2D image, we consider different possible gemetrical structures seen among PNe: mainly bipolar and toroidal shells, as accounted in the previous studies discussed earlier. We assume various combination of a multiple bipolar shells of varying dimensions and orientations. We consider both open and closed lobe geometry for the bipolar shells. We vary density of the shell components as well as their associated velocity fields. Finally, the basic 3D mesh structure is best-constructed using three components: two bipolar shells and a toroidal shell surrounding the bipolar shells at their waist. By fitting the PV diagrams (Fig. \ref{fig:pvdiagrams}) along with the nebular image, we attempt to constrain the 3D model by the physical paramters: velocity field; extent of the nebular outflows along the line-of-sight, given by axes ratios of the components; and inclination ($i$) of the nebular structure with the line-of-sight.

\subsubsection{Results from the mopho-kinematic model}

We obtain a morpho-kinematic model of NGC 6572 that best fits our observational data through visual estimation. We find that the bipolar shells are open and closed lobed and have axes ratios $3:1$ and $5.5:1$, respectively. The toroidal shell has $\mathrm{radius:height}=1:3$, and surrounds the bipolar shells at the waist region. We notice that the toroidal axis and the major axes of both the bipolar shells are aligned with each other closer to the waist of the overall structure. However, the lobes gradually get separated as the they move towards the polar regions as they develop a spirality in their ouflows. We obtain the expansion velocity of the nebular outflows: an expansion velocity field proportional to the radial distance from the central star ($V(r)\sim{r}$) satisfactorily fits the observations. The maximum expansion velocity is estimated as $\mathrm{V}_{max}=50$ km s$^{-1}$ (for both bipolar outflows) and $\mathrm{V}_{max}=20$ km s$^{-1}$ (for the toroidal shell). In this work, we estimate $i=40^{\circ}$, which matches very well with that obtained by \citet{1994MNRAS.269..975H}. Since we use only two slit position angles to fit our observations, there might be certain amount of degeneracy in the estimation of the parameters. For example, theoretically, higher $i$ and lower axes ratio, or, lower $i$ and higher axes ratio could have produced the observed image as well. However, we test our model by varying the parameters in both higher and lower ranges. From visual inspection of the deviation of the model image from the observed image, we estimate an error of 15-20$\%$ in the model parameters.   

\begin{table*}
\centering
\small
\caption{{ Observed extincted (F($\lambda$)) and dereddened (I($\lambda$)) emission line fluxes for the EW and NS spectra.} \label{tab:obsflux}}
\begin{tabular}{cccccc}
\hline
 \multicolumn{6}{l}{$c(\mathrm{H}\beta)=0.14\pm0.05$}\\
\hline
 Wavelength {\AA} & Ion &\multicolumn{2}{c}{EW spectral fluxes} & \multicolumn{2}{c}{NS spectral fluxes}\\
& & F($\lambda$) & I($\lambda$) & F($\lambda$) & I($\lambda$) \\ 
\hline
F, I(H$\beta$) & & $1.18\times10^{-10}$ &	$1.63\times10^{-10}$ & $1.12\times10^{-10}$ & $1.55\times10^{-10}$	\\
 (erg cm$^{-2}$ s$^{-1}$) &&&&&\\ 
\hline
\multicolumn{6}{c}{Fluxes normalized with respect to F, I(H$\beta$)=100}\\
&&&&\\
{	3727	{\AA}	}	&	{	$\mathrm{[O~{\sc II}]}$		}	&	{	$	42.421		\pm	0.746$	}	&	{	$	47.120 		\pm	1.937$	}	&	{	$	42.187		\pm	1.142$	}	&	{	$	46.860 		\pm	2.154	$	}	\\
{	3869	{\AA}	}	&	{	$\mathrm{[Ne~{\sc III}]}$	}	&	{	$	136.976		\pm	4.716$	}	&	{	$	150.617 	\pm	7.244$	}	&	{	$	105.563		\pm	4.859$	}	&	{	$	116.076 	\pm	6.614  	$	}	\\
{	4363	{\AA}	}	&	{	$\mathrm{[O~{\sc III}]}$	}	&	{	$	8.724		\pm	0.137$	}	&	{	$	9.160 		\pm	0.214$	}	&	{	$	7.932		\pm	0.252$	}	&	{	$	8.328 		\pm	0.301  	$	}	\\
{	4471	{\AA}	}	&	{	$\mathrm{He~{\sc I}}$		}	&	{	$	4.642		\pm	0.197$	}	&	{	$	4.820 		\pm	0.214$	}	&	{	$	4.890		\pm	0.135$	}	&	{	$	5.077 		\pm	0.156  	$	}	\\
{	4686	{\AA}	}	&	{	$\mathrm{He~{\sc II}}$		}	&	{	$	0.666		\pm	0.009$	}	&	{	$	0.677 		\pm	0.010$	}	&	{	$	0.715		\pm	0.030$	}	&	{	$	0.727 		\pm	0.031  	$	}	\\
{	4711	{\AA}	}	&	{	$\mathrm{[Ar~{\sc IV}]}$	}	&	{	$	1.360		\pm	0.043$	}	&	{	$	1.379 		\pm	0.044$	}	&	{	$	1.266		\pm	0.066$	}	&	{	$	1.284 		\pm	0.067  	$	}	\\
{	4740	{\AA}	}	&	{	$\mathrm{[Ar~{\sc IV}]}$	}	&	{	$	1.644		\pm	0.101$	}	&	{	$	1.662 		\pm	0.102$	}	&	{	$	1.772		\pm	0.099$	}	&	{	$	1.792 		\pm	0.100  	$	}	\\
{	4861	{\AA}	}	&	{	$\mathrm{H~{\sc I}}$		}	&	{	$	100.000		\pm	3.392$	}	&	{	$	100.000 	\pm	3.392$	}	&	{	$	100.000		\pm	6.540$	}	&	{	$	100.000 	\pm	6.540  	$	}	\\
{	4959	{\AA}	}	&	{	$\mathrm{[O~{\sc III}]}$	}	&	{	$	407.136		\pm	8.566$	}	&	{	$	403.729 	\pm	8.579$	}	&	{	$	397.702		\pm	11.102$	}	&	{	$	394.374 	\pm	11.071  $	}	\\
{	5007	{\AA}	}	&	{	$\mathrm{[O~{\sc III}]}$	}	&	{	$	1211.424	\pm	30.314$	}	&	{	$	1196.566 	\pm	30.394$	}	&	{	$	1192.153	\pm	35.109$	}	&	{	$	1177.531 	\pm	35.057  $	}	\\
{	5518	{\AA}	}	&	{	$\mathrm{[Cl~{\sc III}]}$	}	&	{	$	0.259		\pm	0.025$	}	&	{	$	0.247 		\pm	0.024$	}	&	{	$	0.189		\pm	0.005$	}	&	{	$	0.180 		\pm	0.006  	$	}	\\
{	5538	{\AA}	}	&	{	$\mathrm{[Cl~{\sc III}]}$	}	&	{	$	0.440		\pm	0.012$	}	&	{	$	0.419 		\pm	0.014$	}	&	{	$	0.378		\pm	0.011$	}	&	{	$	0.360 		\pm	0.012  	$	}	\\
{	5755	{\AA}	}	&	{	$\mathrm{[N~{\sc II}]}$		}	&	{	$	1.726		\pm	0.059$	}	&	{	$	1.625 		\pm	0.065$	}	&	{	$	2.116		\pm	0.067$	}	&	{	$	1.992 		\pm	0.076  	$	}	\\
{	5876	{\AA}	}	&	{	$\mathrm{He~{\sc I}}$		}	&	{	$	11.973		\pm	0.908$	}	&	{	$	11.205 		\pm	0.889$	}	&	{	$	14.802		\pm	0.632$	}	&	{	$	13.853 		\pm	0.675  	$	}	\\
{	6300	{\AA}	}	&	{	$\mathrm{[O~{\sc I}]}$		}	&	{	$	5.008		\pm	0.155$	}	&	{	$	4.597 		\pm	0.199$	}	&	{	$	6.640		\pm	0.193$	}	&	{	$	6.095 		\pm	0.256  	$	}	\\
{	6312	{\AA}	}	&	{	$\mathrm{[S~{\sc III}]}$	}	&	{	$	0.967		\pm	0.093$	}	&	{	$	0.887 		\pm	0.090$	}	&	{	$	0.963		\pm	0.095$	}	&	{	$	0.883 		\pm	0.091  	$	}	\\
{	6364	{\AA}	}	&	{	$\mathrm{[O~{\sc I}]}$		}	&	{	$	1.652		\pm	0.046$	}	&	{	$	1.512 		\pm	0.063$	}	&	{	$	2.235		\pm	0.056$	}	&	{	$	2.046 		\pm	0.082  	$	}	\\
{	6548	{\AA}	}	&	{	$\mathrm{[N~{\sc II}]}$		}	&	{	$	17.476		\pm	1.932$	}	&	{	$	15.869 		\pm	1.836$	}	&	{	$	21.496		\pm	1.609$	}	&	{	$	19.520 		\pm	1.606  	$	}	\\
{	6563	{\AA}	}	&	{	$\mathrm{H~{\sc I}}$		}	&	{	$	276.480		\pm	5.679$	}	&	{	$	250.896 	\pm	10.041$	}	&	{	$	346.163		\pm	7.291$	}	&	{	$	314.131 	\pm	12.657  $	}	\\
{	6584	{\AA}	}	&	{	$\mathrm{[N~{\sc II}]}$		}	&	{	$	66.557		\pm	1.500$	}	&	{	$	60.343 		\pm	2.495$	}	&	{	$	81.131		\pm	1.771$	}	&	{	$	73.557 		\pm	3.014  	$	}	\\
{	6678	{\AA}	}	&	{	$\mathrm{He~{\sc I}}$		}	&	{	$	3.238		\pm	0.065$	}	&	{	$	2.924 		\pm	0.121$	}	&	{	$	4.792		\pm	0.089$	}	&	{	$	4.327 		\pm	0.176  	$	}	\\
{	6716	{\AA}	}	&	{	$\mathrm{[S~{\sc II}]}$		}	&	{	$	0.720		\pm	0.023$	}	&	{	$	0.649 		\pm	0.032$	}	&	{	$	1.199		\pm	0.035$	}	&	{	$	1.081 		\pm	0.051  	$	}	\\
{	6731	{\AA}	}	&	{	$\mathrm{[S~{\sc II}]}$		}	&	{	$	1.637		\pm	0.034$	}	&	{	$	1.475 		\pm	0.062$	}	&	{	$	2.462		\pm	0.034$	}	&	{	$	2.218 		\pm	0.087  	$	}	\\
{	7065	{\AA}	}	&	{	$\mathrm{He~{\sc I}}$		}	&	{	$	8.373		\pm	0.168$	}	&	{	$	7.435 		\pm	0.346$	}	&	{	$	12.131		\pm	0.192$	}	&	{	$	10.771 		\pm	0.484  	$	}	\\
{	7136	{\AA}	}	&	{	$\mathrm{[Ar~{\sc III}]}$	}	&	{	$	14.670		\pm	0.279$	}	&	{	$	12.986 		\pm	0.612$	}	&	{	$	20.395		\pm	0.330$	}	&	{	$	18.053 		\pm	0.832  	$	}	\\
{	7325	{\AA}	}	&	{	$\mathrm{[O~{\sc II}]}$		}	&	{	$	12.045		\pm	0.826$	}	&	{	$	10.574 		\pm	0.874$	}	&	{	$	15.327		\pm	0.402$	}	&	{	$	13.456 		\pm	0.713  	$	}	\\
{	7751	{\AA}	}	&	{	$\mathrm{[Ar~{\sc III}]}$	}	&	{	$	3.260		\pm	0.082$	}	&	{	$	2.810 		\pm	0.164$	}	&	{	$	4.694		\pm	0.086$	}	&	{	$	4.046 		\pm	0.225  	$	}	\\
{	8046	{\AA}	}	&	{	$\mathrm{[Cl~{\sc IV}]}$	}	&	{	$	0.271		\pm	0.008$	}	&	{	$	0.231 		\pm	0.015$	}	&	{	$	0.420		\pm	0.008$	}	&	{	$	0.358 		\pm	0.021  	$	}	\\
{	9069	{\AA}	}	&	{	$\mathrm{[S~{\sc III}]}$	}	&	{	$	15.162		\pm	0.993$	}	&	{	$	12.492 		\pm	1.184$	}	&	{	$	21.528		\pm	1.551$	}	&	{	$	17.737 		\pm	1.763  	$	}	\\
\hline
\end{tabular}
\end{table*} 

\subsection{Spectral emission line analyses}

\subsubsection{Emission line fluxes}
 
\begin{figure}
\centering
\scalebox{0.45}[0.45]{\includegraphics{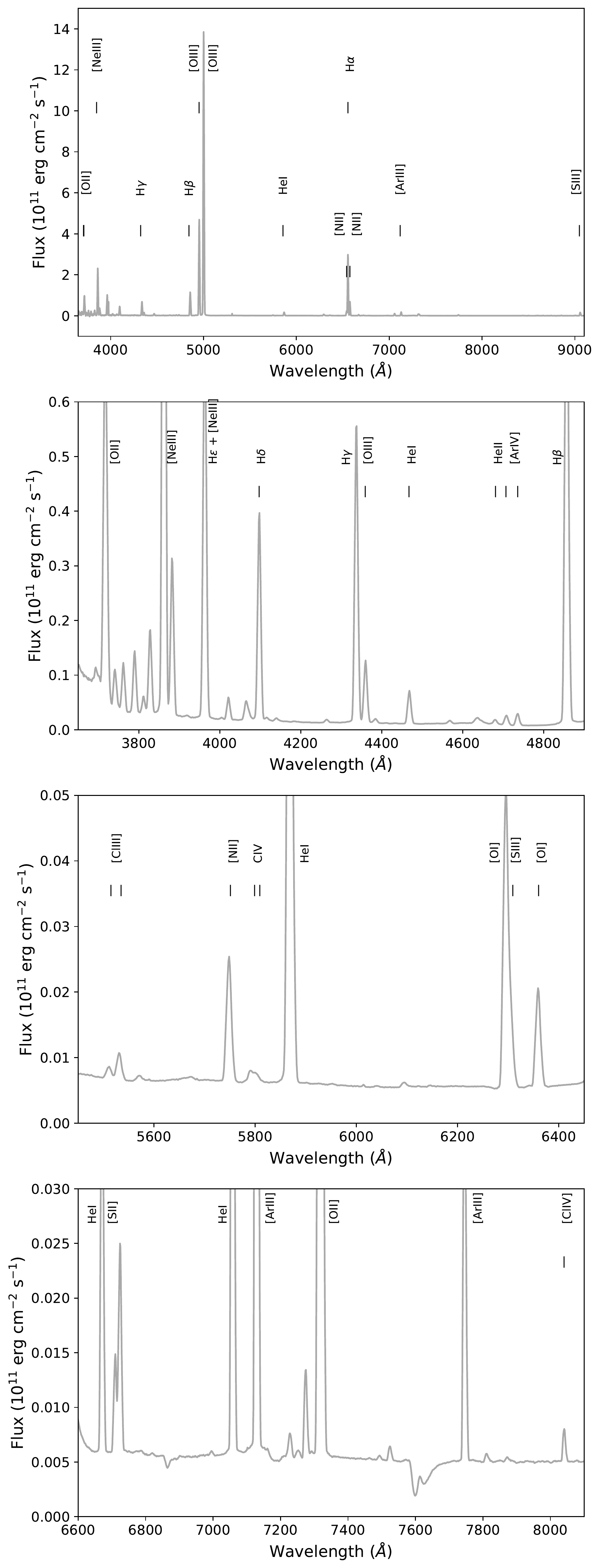}}
 \caption{{ The EW spectra of NGC 6572 in different spectral ranges across the optical region.} \label{fig:specoptew}}
\end{figure}

\begin{figure}
\centering
\scalebox{0.45}[0.45]{\includegraphics{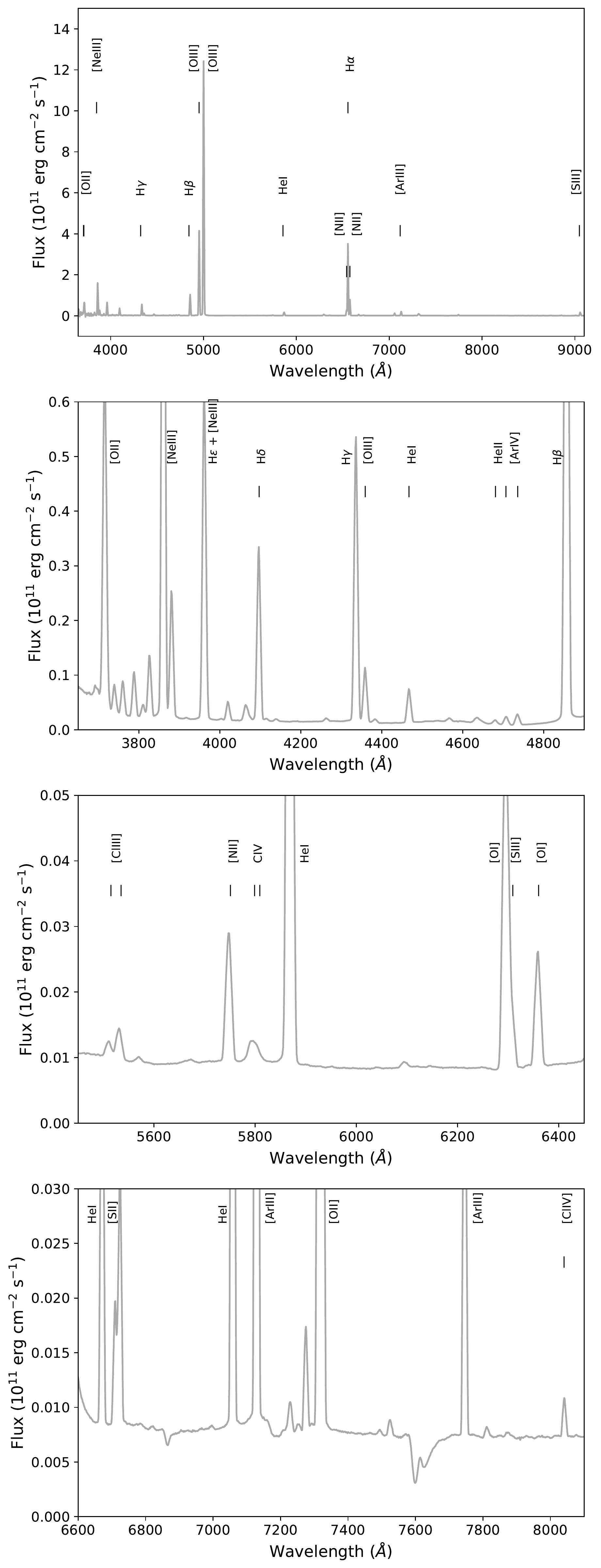}} 
\caption{{ The NS spectra of NGC 6572 in different spectral ranges across the optical region.} \label{fig:specoptns}}
\end{figure}

The optical spectra of NGC 6572 shows strong emission line features and gives the signatures of a moderately ionized PN. We used the \textsc{specutils} package \citep{nicholas_earl_2022_6207491} to fit the emission lines using single gaussian profiles and estiamte their fluxes.
To calculate the errors in the flux measurements, we used the expression given by \cite{Lenz_1992} as 
\begin{equation}
\frac{F}{{\sigma}_F} = 0.67\left(\frac{FWHM}{{\Delta}\lambda}\right)^{0.5}\frac{f_0}{{\sigma}_0}.
\end{equation}
$F$ is the calculated flux and ${\sigma}_F$ is the error in flux. FWHM refers to the full width at half-maximum of the emission lines. $f_0$ is the peak intensity of any emission line and ${\sigma}_0$ is the noise level at the peak position. The extincted and the dereddened (see below) emission line fluxes for the EW and NS spectra are shown in Table \ref{tab:obsflux}. The dereddened EW and NS spectra are shown in Fig. \ref{fig:specoptew} and Fig. \ref{fig:specoptns}, respectively.

From our HCT spectra, we measure the the logarithmic extinction at H$\beta$ ($c(\mathrm{H}\beta)$) for NGC 6572 and apply interstellar extinction correction to the observed spectra using the relation,
\begin{equation}
I(\lambda)=F(\lambda)10^{[c(\mathrm{H}\beta)f(\lambda)]}.
\end{equation}
Here, $I(\lambda)$ and $F(\lambda)$ measures the intrinsic and observed fluxes, respectively. We use the extinction function ($f(\lambda)$) given by \cite{1989ApJ...345..245C}. The value of $c(\mathrm{H}\beta)$ is calculated by equating the observed Balmer line ratios to their theoretical values at electron temperature, $T_\mathrm{e}=10^4$ K and electron density, $N_\mathrm{e}=10^4$ cm$^{-3}$ \citep{2006agna.book.....O}.
 
From our HCT observations made on 2021 May 29, we obtain $c(\mathrm{H}\beta)\sim0.00$, i.e., nearly a zero-extinction (see discussion below), for the EW spectra,     
whereas, for the NS spectra we have $c(\mathrm{H}\beta)=0.28\pm0.1$. Finally, we measure the amount of interstellar extinction towards NGC 6572 as the average of $c(\mathrm{H}\beta)$ values obtained from the EW and NS spectra. Hence, we adopt $c(\mathrm{H}\beta)=0.14\pm0.05$ to correct our observed HCT spectra for interstellar extinction.

Few of the earlier estimations of $c(\mathrm{H}\beta)$ are: 0.34 \citep{1992A&AS...94..399C}; 0.07 and 0.73 (\citealt{1992A&AS...95..337T}, from two different observations); 0.48 \citep{2004MNRAS.353.1231L}; 0.30 \citep{2007A&A...463..265G}; 0.41 \citep{2013MNRAS.431....2F}. Since, none of the previous measurements report $c(\mathrm{H}\beta)\sim0.0$, we took another observation (see Table \ref{tab:logobs}). This time we obtain $c(\mathrm{H}\beta)=0.06\pm0.007$ for the EW spectra and $c(\mathrm{H}\beta)=0.29\pm0.018$ for NS spectra. Hence, $c(\mathrm{H}\beta)$ measured for EW spectra slightly increases with time, however, still remains lower than that from the NS spectra.

The difference in the estimations of $c(\mathrm{H}\beta)$ discussed above might be due to the difference in the slit dimesions and orientations in different observations, which led to difference in the extraction regions. \citet{2016MNRAS.455..930A} finds $c(\mathrm{H}\beta)$ to vary from $0.24\pm0.05$ to $0.29\pm0.06$ { along the nebula}. However, our measurements show a larger variation within the nebula. Moreover, the reported value of $c(\mathrm{H}\beta)=0.48$ by \citet{2004MNRAS.353.1231L} were form observations taken with EW slit orientation, in large contrast to our estimations. Hence, there could also be a time variation of $c(\mathrm{H}\beta)$ estimated for NGC 6572, particularly from the observations of waist region. Previously, time variation of $c(\mathrm{H}\beta)$ for SaSt 2-3 was reported \citep{2019MNRAS.482.2354O}. Further investigations in this regard might identify the reason for reportedly large variation of $c(\mathrm{H}\beta)$ measured for NGC 6572.

While, we observe a few differences between the line fluxes of the EW and the NS spectra, overall characteristics seem not to differ much. The EW spectrum corresponds the central waist region (close to the central star) of the nebulae, hence, show a slightly higher ionization than depicted through the NS spectrum, which partially accounts for the waist and mostly comes from the polar lobes. 

\begin{figure*}
\centering
\scalebox{0.5}[0.5]{\includegraphics{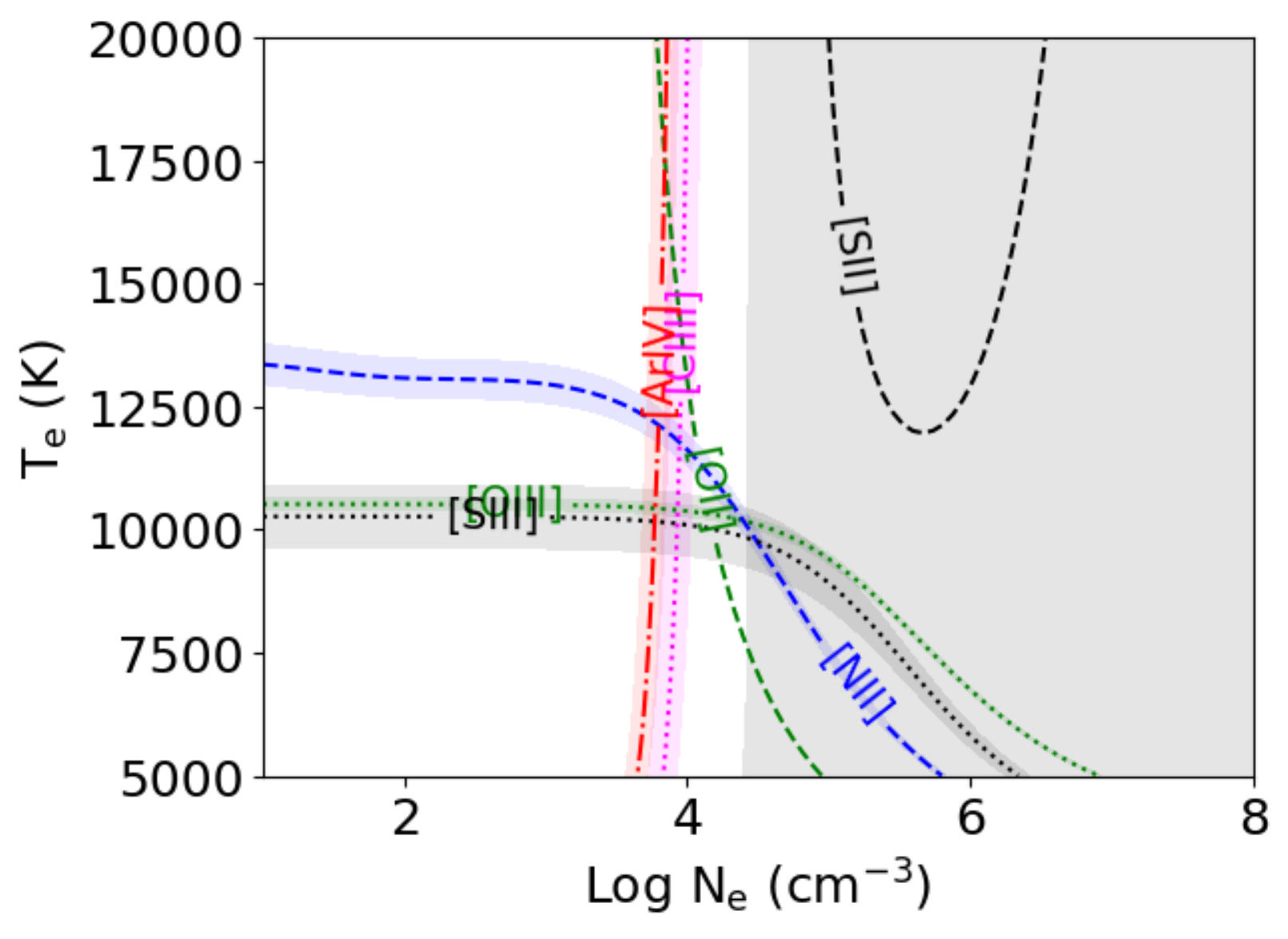}}
\scalebox{0.5}[0.5]{\includegraphics{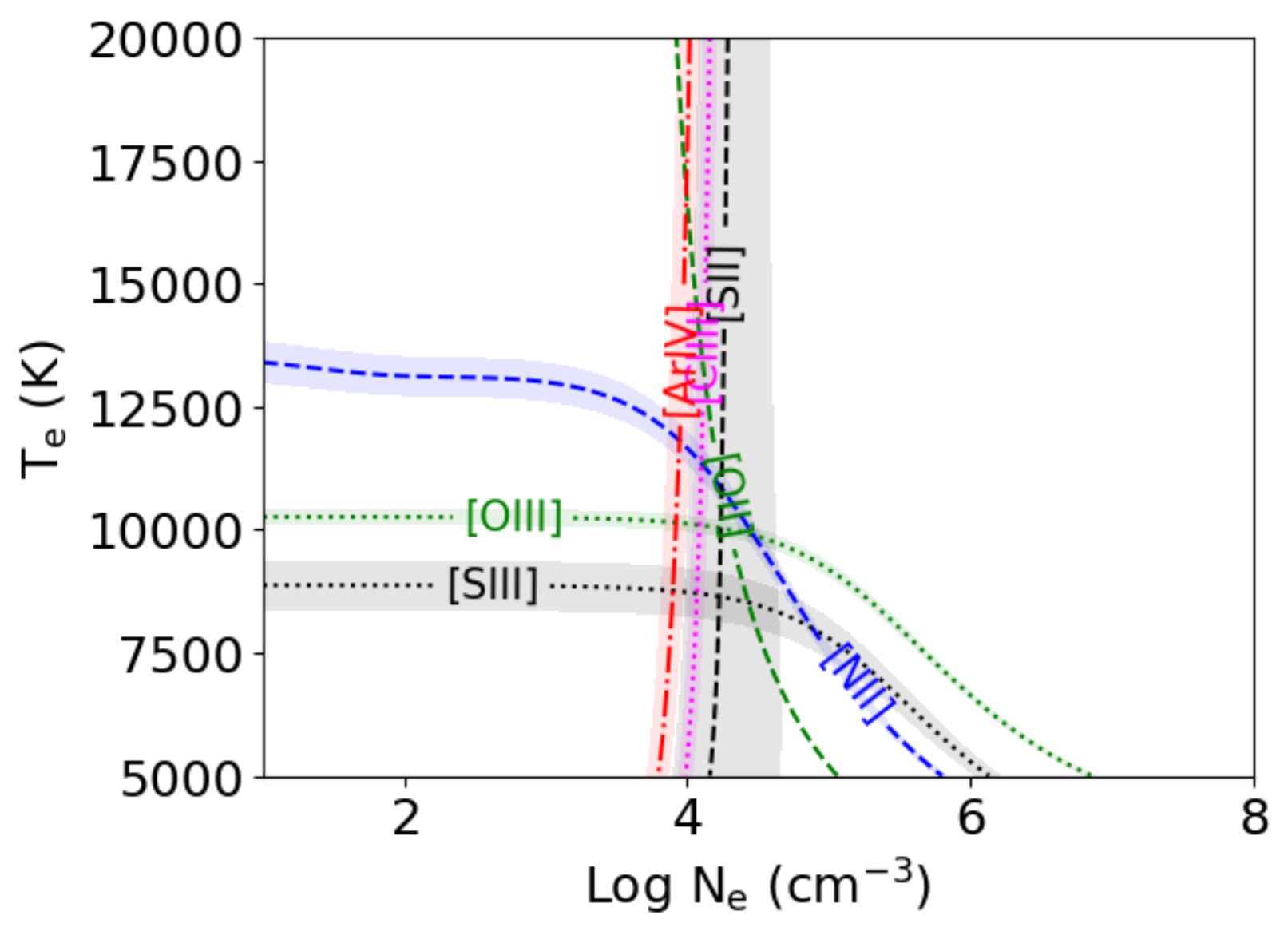}}
 \caption{The $T_\mathrm{e}$ vs Log $N_\mathrm{e}$ plot for NGC 6572 for two slit positions EW (left) and NS (right). The labels with the graphs denote the corresponding ion used for generating the same. \label{fig:tenengc6572}}
\end{figure*}

\begin{table*}
\centering
\small
\caption{$T_\mathrm{e}$ and $N_\mathrm{e}$ calculated from EW and NS spectra of NGC 6572. \label{tab:tenengc6572}}
 \begin{tabular}{lcclcc}
 \hline
$T_\mathrm{e}$ diagnostic line ratio & Ratio & $T_\mathrm{e}$ & $N_\mathrm{e}$ diagnostic line ratio & Ratio & $N_\mathrm{e}$\\
 \hline
\multicolumn{6}{c}{Slit-orientation EW}\\
{ $T_\mathrm{e}$([O~{\sc iii}] 4363/5007)} & { 0.0076} & { ${10391\pm43}$} & { $N_\mathrm{e}$([Ar~{\sc iv}] 4740/4711)} & { 1.21} & { ${5097\pm422}$}\\
{ $T_\mathrm{e}$([S~{\sc iii}] 6312/9069)} & { 0.071} & { ${10047\pm318}$} & { $N_\mathrm{e}$([Cl~{\sc iii}] 5538/5518)} & { 1.70} & { ${9984\pm2495}$}\\
{ $T_\mathrm{e}$([N~{\sc ii}] 5755/6584)} & { 0.027} & { ${11752\pm332}$} & { $N_\mathrm{e}$([Cl~{\sc iii}] 5538/5518)} & { 1.70} & { ${10398\pm2630}$}\\
{ $T_\mathrm{e}$([N~{\sc ii}] 5755/6584)} & { 0.027} & { ${10041\pm243}^a$} & { $N_\mathrm{e}$([S~{\sc ii}] 6731/6716)} & { 2.27} & { ${>26670}$}\\
\hline
\multicolumn{6}{c}{Slit-orientation NS}\\
{ $T_\mathrm{e}$([O~{\sc iii}] 4363/5007)} & { 0.0071} & { ${10175\pm116}$} & { $N_\mathrm{e}$([Ar~{\sc iv}] 4740/4711)} & { 1.40} & { ${8925\pm771}$}\\
{ $T_\mathrm{e}$([S~{\sc iii}] 6312/9069)} & { 0.050} & { ${8569\pm190}$} & { $N_\mathrm{e}$([Cl~{\sc iii}] 5538/5518)} & { 2.00} & { ${11474\pm1053}$}\\
{ $T_\mathrm{e}$([N~{\sc ii}] 5755/6584)} & { 0.027} & { ${10820\pm280}$} & { $N_\mathrm{e}$([S~{\sc ii}] 6731/6716)} & { 2.05} & { ${17742\pm7295}$}\\
 \hline
\multicolumn{6}{l}{ $^a$Using lower limit of $N_\mathrm{e}$([S~{\sc ii}])}\\
 \end{tabular}
\end{table*}

\begin{table*}
\centering
\small
\caption{Nebular ionic and total abundances of NGC 6572 { calcualted directly from emission line fluxes} ($A(B)=A\times10^B$). \label{tab:abunngc6572}}
\begin{tabular}{l c c c c c}
\hline
X & X$^{+}$/H & X$^{2+}$/H & X$^{3+}$/H & ICF & $12+\mathrm{X/H}$\\
\hline
\multicolumn{6}{c}{Slit-orientation EW}\\
&&&&&\\
He	&	{ $0.068 \pm 0.006$}	&	{ $\left(5.68 \pm 0.09\right) \times 10^{-4}$}	 	&	-		& 1.0 					& { $10.84\pm0.04$}\\
C	&	-		&	-		&	-	    & - 		& - \\
N	&	{ $\left(1.07 \pm 0.04\right) \times 10^{-5}$}	&	-		&	-							& { $13.46\pm1.07$}	& { $8.17\pm0.04$}\\
O	&	{ $\left(3.61 \pm 0.29\right) \times 10^{-5}$}		&	{ $\left(3.04 \pm 0.07\right) \times 10^{-4}$}	&	-		& { $1.00\pm0.001$} 	& { $8.534\pm0.01$}\\
Ne	&	-		&	{ $\left(1.12 \pm 0.05\right) \times 10^{-4}$}	&	-							& { $1.22\pm0.01$}	& { $8.144\pm0.02$}\\
S	&	{ $\left(1.18 \pm 0.06\right) \times 10^{-7}$}	&	{ $\left(1.27 \pm 0.13\right) \times 10^{-6}$}&	-				& { $1.48\pm0.04$} 	& { $6.609\pm0.014$}\\
Cl	&	-		&	{ $\left(3.78 \pm 0.12\right) \times 10^{-8}$}	&	{ $\left(1.31 \pm 0.08\right) \times 10^{-8}$}		& { $1.63\pm0.03$} 	& { $5.049\pm0.008$}\\
Ar	&	-		&	{ $\left(8.6 \pm 0.4\right) \times 10^{-7}$}	&	{ $\left(3.11 \pm 0.19\right) \times 10^{-7}$}		& { $1.42\pm0.02$} 	& { $6.258\pm0.014$}\\
\hline
\multicolumn{6}{c}{Slit-orientation NS}\\
&&&&&\\
He	&	{ $0.085 \pm 0.004$}	&	{ $\left(6.11 \pm 0.26\right) \times 10^{-4}$}	 	&	-		& 1.0 					& { $10.932\pm0.021$}\\
C	&	-		&	-		&	-	    & - 		& - \\
N	&	{ $\left(1.31 \pm 0.05\right) \times 10^{-5}$}	&	-		&	-							& { $10.91\pm0.63$} 	& { $8.163\pm0.029$}\\
O	&	{ $\left(4.57 \pm 0.25\right) \times 10^{-5}$}	&	{ $\left(3.00 \pm 0.09\right) \times 10^{-4}$}	&	-			& { $1.00\pm0.001$} 	& { $8.541\pm0.012$}\\
Ne	&	-		&	{ $\left(8.6 \pm 0.5\right) \times 10^{-5}$}	&	-							& { $1.25\pm0.01$} 	& { $8.052\pm0.023$}\\
S	&	{ $\left(1.56 \pm 0.07\right) \times 10^{-7}$}	&	{ $\left(1.26 \pm 0.13\right) \times 10^{-6}$}&	-				& { $1.37\pm 0.03$}	& { $6.603\pm0.013$}\\
Cl	&	-		&	{ $\left(3.24 \pm 0.11\right) \times 10^{-8}$}	&	{ $\left(2.03 \pm 0.11\right) \times 10^{-8}$}		& { $1.56\pm0.02$}	& { $5.046\pm0.008$}\\
Ar	&	-		&	{ $\left(1.20 \pm 0.06\right) \times 10^{-6}$}		&	{ $\left(3.34 \pm 0.18\right) \times 10^{-7}$}		& { $1.37\pm0.01$} 	& { $6.342\pm0.015$}\\
\hline
\end{tabular}
\end{table*}

\subsubsection{Plasma parameters and nebular abundances} \label{sec:nebular}

We obtain the characteristic plasma parameters, $T_\mathrm{e}$ and $N_\mathrm{e}$, using the corresponding characteristic line ratios obtained from our HCT spectra. We use the package \textsc{pyneb} \citep{2015A&A...573A..42L} to calculate the quantities. We obtain $T_\mathrm{e}$ from the line ratios [O~{\sc iii}] 4363/5007 {\AA},  [N~{\sc ii}] 5755/6584, and [S~{\sc iii}] 6312/9069 {\AA}. [Ar~{\sc iv}] 4740/4711 {\AA}, [Cl~{\sc iii}] 5538/5518 {\AA}, and [S~{\sc ii}] 6731/6716 {\AA} are used for $N_\mathrm{e}$ calculations. Table \ref{tab:tenengc6572} lists the $T_\mathrm{e}$ and $N_\mathrm{e}$ values obtained for the EW and NS spectra and corresponding line flux ratios that are used for the calculations. Fig. \ref{fig:tenengc6572} shows $T_\mathrm{e}$-$N_\mathrm{e}$ plots obtained for the EW and NS spectra. From the $T_\mathrm{e}$-$N_\mathrm{e}$ plots, we obtain approximated values of $T_\mathrm{e}\sim11,500$ K and $N_\mathrm{e}\sim10000$ cm$^{-3}$ for the calculation of ionic abundances directly from the emission line fluxes (Table \ref{tab:abunngc6572}). These are also separately obtained for the EW and NS spectra. From the ionic abundances, we obtain the total elemental abundances (Table \ref{tab:abunngc6572}). We apply the ionization correction factors (ICFs) calculated using the formulae given by \citet{2014MNRAS.440..536D}.

\subsection{Photoionization modelling}

\subsubsection{Basic method of modelling}

We study the ionization structure of the collimated ionized shells of NGC 6572 through a photoionization modelling approach. We use the photoionization code \textsc{cloudy} \citep{2017RMxAA..53..385F} and the \textsc{pycloudy} \citep{2013ascl.soft04020M} library to compute the models. In general, a set of input physical parameters, which characterizes the central star and the nebula, are specified as input of the model. \textsc{cloudy} internally solves the radiation transfer, ionization balance and energy balance equations at each point of the nebula, and calculate the model spectrum. A total model spectrum consists of the attenuated spectrum of the central star, the nebular continuum and the emission lines. We aim to generate a total model spectrum that well-reproduces the the observed flux, and the line ratios. The specific observables we fit in our modelling process include the $\mathrm{H}\beta$ received on Earth and the line flux ratios. Distance to the PN ($d$) is used as a model parameter to calculate line fluxes as on Earth, thus $d$ can be estimated self-consistently from the model. Since the nebular ionization balance is characterized by the line ratios corresponding to the different ionization states of a specific element and the line ratios related to $T_\mathrm{e}$ and $N_\mathrm{e}$, we especially focus on fitting these line ratios properly. \textsc{pycloudy} allows the user to take into account the slit dimensions and extraction windows used during actual observations and data reduction, respectively. Also, the nebular inclination with the observers line-of-sight can be applied in the \textsc{pycloudy} models. Hence, we compare our observed fluxes, extracted as shown in Fig. \ref{fig:ngc6572} with the model fluxes obtained using a synthetic extraction window during our model calculations.      

\begin{figure}
\centering
\scalebox{0.4}[0.4]{\includegraphics{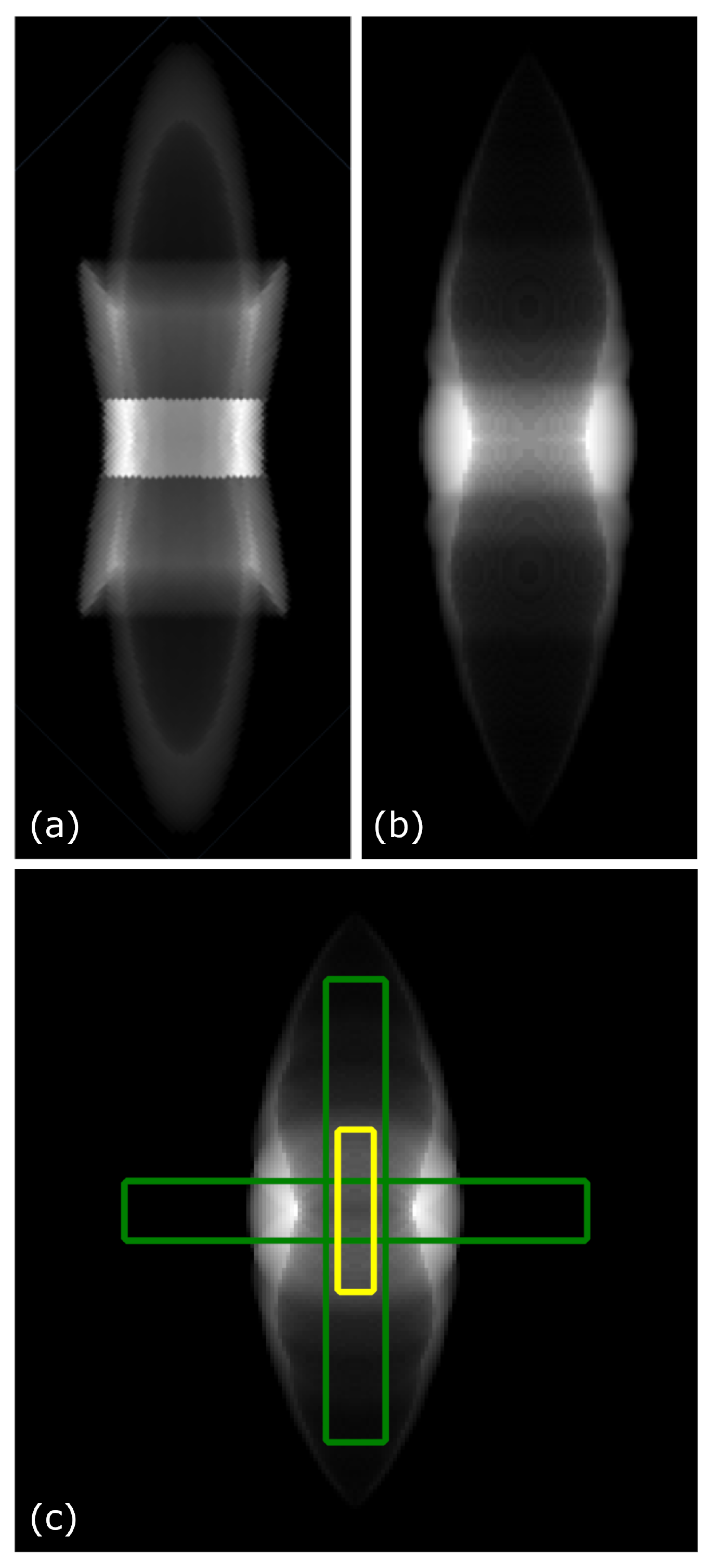}}
 \caption{(a) The simplified 3D structure of NGC 6572 derived from the actual 3D model to use as reference of the {\sc pycloudy} model (b) The modelled photoionized shell of NGC 6572 generated using {\sc pycloudy}. (c) The modelled photoionized nebula, inclined at $i=40^{\circ}$, and superposed synthetic with slits replicating the HCT (green) and WHT (yellow) apertures through which the modelled fluxes (Table \ref{tab:obsvsmod}) and modelled line-profiles (Fig. \ref{fig:WHTprofiles}) are obtained, respectively. \label{fig:shapepycloudy}}
\end{figure}

\begin{table*}
\centering
\small
\caption{{ Physical parameters of the photoionization model} of NGC 6572. \label{tab:pycloudymodel}}
\begin{tabular}{l c c c}
\hline
{ $d$ $(\mathrm{kpc})$} &&& 1.85\\
$T_\mathrm{eff}$ $(\mathrm{K})$ &&& 68000\\
$L$ ($L_{\sun}$) &&& { 5700}\\
Log $g$ $(\mathrm{cm}$ $\mathrm{s^{-2}})$ &&& 5.0\\
Log $Z$ &&& 0.0\\
$f$ &&& 0.5\\
$M_{H^{+}}$ ($M_{\sun}$) &&& { 0.047}\\
$M_{H^{0}}$ ($M_{\sun}$) &&& { 0.004}\\
$M_{H}$ ($M_{\sun}$) &&& { 0.051}\\
Shell Geometry &&& Bipolar\\
\multicolumn{4}{l}{Shell configuration}\\
$\theta$ ($^{\circ}$) & Log $r_\mathrm{in}$ $(\mathrm{cm})$ & Log $r_\mathrm{out}$ $(\mathrm{cm})$ & Log $n_\mathrm{H}$ $(\mathrm{cm^{-3}})$\\
\hline
0 & 16.68 & 16.97 & 4.51 \\
10 & 16.71 & 16.98 & 4.51 \\
20 & 16.74 & 16.98 & 4.51 \\
30 & 16.81 & 17.00 & 4.51 \\
40 & 16.86 & 17.08 & 4.38 \\
50 & 16.99 & 17.12 & 4.38 \\
60 & 17.14 & 17.20 & 4.38 \\
70 & 17.23 & 17.29 & 4.23 \\
80 & 17.38 & 17.41 & 4.23 \\
90 & 17.52 & 17.53 & 4.23 \\
\hline
\multicolumn{4}{l}{Elemental abundances ($12+\mathrm{Log\:(X/H)}$, $\mathrm{X=element}$)}\\
\multicolumn{4}{l}{He: 10.85, C: 9.00, N: 8.08, O: 8.65}\\
\multicolumn{4}{l}{Ne: 8.27, S: 6.50, Cl: 4.86, Ar: 6.23}\\
\hline
\end{tabular}
\end{table*} 

\begin{table}
\centering
\small
\caption{Comparison between observed and modelled line fluxes of NGC 6572. 
\label{tab:obsvsmod}}
\begin{tabular}{lcccc}
\hline
{ Emission lines} & \multicolumn{2}{c}{Observed} & \multicolumn{2}{c}{Modelled} \\
&\multicolumn{2}{c}{flux} & \multicolumn{2}{c}{flux}  \\
& EW & NS & EW & NS\\
\hline
Log $I(\mathrm{H}\beta)$	& 	$ -9.79$	&	$ -9.81$	&	$ -9.85$ & $ -9.97$	\\
(erg cm$^{-2}$ s$^{-1}$)	&	&	& &	\\
\hline
\multicolumn{4}{c}{Fluxes normalized with respect to I(H$\beta$)=100}\\
&&&\\
	H$\beta$					4861	{\AA}	&	100.000	&	100.000 	&	100.00		&	 100.00		\\
	H$\alpha$					6563	{\AA}	&	250.90 	&	314.13	 	&	272.82		&	 272.83		\\
$\mathrm{He~{\sc I}		}$		5876	{\AA}	&	11.21 	&	13.85 		&	 15.23		&	  14.89		\\
$\mathrm{He~{\sc I}		}$		6678	{\AA}	&	2.92 	&	4.33 		&	  4.14		&	   4.04		\\
$\mathrm{He~{\sc I}		}$		7065	{\AA}	&	7.44 	&	10.77 		&	  9.29		&	   8.76		\\
$\mathrm{He~{\sc II}	}$		4686	{\AA}	&	0.68 	&	0.73 		&	  0.06		&	   0.05		\\
$\mathrm{[N~{\sc II}]	}$		5755	{\AA}	&	1.63 	&	1.99 		&	  1.99		&	   2.28		\\
$\mathrm{[N~{\sc II}]	}$		6548	{\AA}	&	15.87 	&	19.52 		&	 20.64		&	  26.04		\\
$\mathrm{[N~{\sc II}]	}$		6583	{\AA}	&	60.34 	&	73.56 		&	 60.86		&	  76.77		\\
$\mathrm{[O~{\sc I}]	}$		6300	{\AA}	&	4.60 	&	6.10 		&	  4.40		&	   5.29		\\
$\mathrm{[O~{\sc I}]	}$		6364	{\AA}	&	1.51 	&	2.05 		&	  1.40		&	   1.69		\\
$\mathrm{[O~{\sc II}]	}$		3727	{\AA}	&	47.12 	&	46.86 		&	 35.56		&	  48.24		\\
$\mathrm{[O~{\sc II}]	}$		7325	{\AA}	&	10.57 	&	13.46 		&	 16.22		&	  18.55		\\
$\mathrm{[O~{\sc III}]	}$		4363	{\AA}	&	9.16 	&	8.33 		&	  9.88		&	   9.19		\\
$\mathrm{[O~{\sc III}]	}$		4959	{\AA}	&	403.73	&	394.37	 	&	402.24		&	 383.31		\\
$\mathrm{[O~{\sc III}]	}$		5007	{\AA}	&	1196.57 &	1177.53 	&	1200.11		&	 1143.65	\\
$\mathrm{[Ne~{\sc III}]	}$		3869	{\AA}	&	150.61	&	116.08	 	&	152.78		&	 148.52		\\
$\mathrm{[S~{\sc II}]	}$		6716	{\AA}	&	0.65 	&	1.08 		&	  0.61		&	   0.84		\\
$\mathrm{[S~{\sc II}]	}$		6731	{\AA}	&	1.48 	&	2.22 		&	  1.34		&	   1.81		\\
$\mathrm{[S~{\sc III}]	}$		6312	{\AA}	&	0.89 	&	0.88 		&	  0.98		&	   0.99		\\
$\mathrm{[S~{\sc III}]	}$		9069	{\AA}	&	12.49 	&	17.74 		&	 13.56		&	  14.13		\\
$\mathrm{[Cl~{\sc III}]	}$		5518	{\AA}	&	0.25 	&	0.18 		&	  0.13		&	   0.14		\\
$\mathrm{[Cl~{\sc III}]	}$		5538	{\AA}	&	0.42 	&	0.36 		&	  0.40		&	   0.42		\\
$\mathrm{[Cl~{\sc IV}]	}$		8046	{\AA}	&	0.23 	&	0.36 		&	  0.29		&	   0.24		\\
$\mathrm{[Ar~{\sc III}]	}$		7136	{\AA}	&	12.99 	&	18.05 		&	 15.18		&	  15.82		\\
$\mathrm{[Ar~{\sc III}]	}$		7751	{\AA}	&	2.81 	&	4.05 		&	  3.60		&	   3.76		\\
$\mathrm{[Ar~{\sc IV}]	}$		4711	{\AA}	&	1.38 	&	1.28 		&	  1.36		&	   1.14		\\
$\mathrm{[Ar~{\sc IV}]	}$		4740	{\AA}	&	1.66 	&	1.79 		&	  1.66		&	   1.37		\\
\hline
\end{tabular}
\end{table}    

\subsubsection{Shell configuration of NGC 6572 for the model} \label{sec:shell}

Since NGC 6572 has a highly collimated bipolar structure, a spherical model might not be adequate to fit all the observables satisfactorily. We use \textsc{pycloudy} to generate non-spherical nebular shell as input for the photoionization models. The shell is symmetric with respect to both major and minor axes. 
 
For the {\sc pycloudy} modeling, we use a bipolar axisymmetric input geometry, simplified from the actual 3D {\sc shape} model (Sec. \ref{sec:3dmkmodel}) of NGC 6572, as shown in Fig. \ref{fig:shapepycloudy}. We define the radial dimension of the input shell as a function of the azimuthal angle $\theta$ such that it traces out a bipolar shell in the same dimensional proportions to the simplified 3D model. Corresponding to a chosen value of $d$, we calculate the inner radius ($r_\mathrm{in}$) and outer radius ($r_\mathrm{out}$) of the input shell for ten different $\theta$ within $0-90^{\circ}$. We take care that the modelled shell, when projected on the sky plane, would match the angular extent of the observed nebula. As $d$ is varied, the input radii are also varied accordingly to fit the nebular extent on the sky plane. In the actual models in this work, we use the same $r_\mathrm{in}$ as calculated above. We run our models up to the zone where the kinetic temperature falls to 4000 K and most of the ionization ceases, since we model the ionized region (optical spectra) of the nebula in this work. Hence, $r_\mathrm{out}$ is specified by the extent of the ionized region of the modelled nebula. 

The input nebular shell density assumes the form $n_\mathrm{H}=n_\mathrm{H}(r,\theta)$, where $n_\mathrm{H}(r)$ corresponds to the radial hydrogen density profile and $n_\mathrm{H}(\theta)$ gives the theta dependence of the hydrogen density. In this work, $n_\mathrm{H}(r)$ is assumed to be constant. $n_\mathrm{H}(\theta)$ has a varying profile from the waist to the polar regions of the nebula, with reference to the 3D model of the nebula. 

\subsubsection{Parameter space of the model}  \label{sec:modelprocess}

The parameter space of the photoionization model include the parameters of ionizing radiation and ionized nebula. Rauch's stellar atmospheric models\footnote{\url{http://astro.uni-tuebingen.de/~rauch/NLTE/H-Ni/}} \citep{2003A&A...403..709R} are adopted into our \textsc{cloudy} model input to characterize the radiation from the ionizing source, the central star, which is parametrized by effective temperature ($T_\mathrm{eff}$), luminosity ($L$), gravity ($g$) and metallicity ($Z$). $T_\mathrm{eff}$ and $L$ are varied as free parameters in the modelling.   
The variation in $g$ have no significant effect in the modelling in this work. We choose Log $g=5.0$ for the central star, which is close to value of Log $g(=5.5)$ for the central star estimated by \citet{1994MNRAS.269..975H}, and left the parameter unchanged throughout the modelling. We use a solar metallicity (Log $Z=0)$ for the central star in the models.     
  
As described in Sec. \ref{sec:shell}, $d$, $r_\mathrm{in}$, and $r_\mathrm{out}$ are connected parameters and hence varied together. In this work, $d$ is not completely a free parameter, since $d$ is a well-estimated parameter found in the literature for NGC 6572. We vary $d$ in a { narrow} range ($1.5-2$ kpc) (see Sec. \ref{sec:constrain}) and the radii are varied accordingly. 
The $n_\mathrm{H}(r,\theta)$ profile is discussed in Sec. \ref{sec:shell}. The range of input values of $n_\mathrm{H}$ used in the modelling is around the estimated range of $N_\mathrm{e}$ (Sec. \ref{sec:nebular}). Since the nebula seems to be clumpy, we assumed a filling factor, $f<1.0$ in our models.     

The initial input elemental abundances (X/H, where, $\mathrm{X=element}$) are those calculated directly using the emission line fluxes (Sec. \ref{sec:nebular}). 
It might be necessary to vary the abundances of the elements during the modelling to match the corresponding line fluxes. Also, increase and decrease of the abundance of an element may act to cool or heat the nebula, respectively, and hence, regulate the relative strengths of the emission lines of different excitation energies. These properties are used during modelling to fine tune emission line ratios from different ionization states to match with their observed values.
As we could not estimate C/H from direct analysis, we use the initial C/H as given in \citet{2005ApJS..157..371R}. Once the relative flux ratios from all the observed ionization states of X are properly reproduced by varying all other parameters, X/H is tuned to match the fluxes of the states. Sometimes, X/H of different elements may be varied to create the proper ionization balance within the nebula. For example, in this work C/H is tuned at final point of the modelling to reproduce the [O~{\sc iii}] 4363/5007 {\AA} flux (also see discussion in \citealt{10.1093/mnras/stab860}).           

\subsubsection{Results from the photionization model} \label{sec:bestmodel}

The model results are summarised in Table \ref{tab:pycloudymodel}. The comparison of the observed EW and NS spectral fluxes with modelled fluxes through synthetic extraction windows oriented along EW and NS, respectively are given in Table \ref{tab:obsvsmod}. From our best-fitting model, we obtain an overall good match of the observed and modelled emission line fluxes. From our modelling we estimate that NGC 6572 have a central star of $T_\mathrm{eff}=68000$ K and $L=5700$ $L_{\sun}$. The detailed shell configuration of the outer shell in the best-fitting model parametrized by $r_\mathrm{in}$, $r_\mathrm{out}$, and $n_\mathrm{H}(\theta)$ values are given in Table \ref{tab:pycloudymodel}, corresponding to $d=1.85$ kpc (see Sec. \ref{sec:constrain}). We qualitatively obtained $f=0.5$ for the nebular shell from our model (see Sec. \ref{sec:constrain}). Further, we obtain the total H mass as $M_H=0.051\:M_{\sun}$, where the ionized and neutral mass is estimated as $M_{H^{+}}=0.047\:M_{\sun}$ and $M_{H^{0}}=0.004\:M_{\sun}$, respectively.

\begin{figure*}
\centering
\scalebox{0.55}[0.55]{\includegraphics{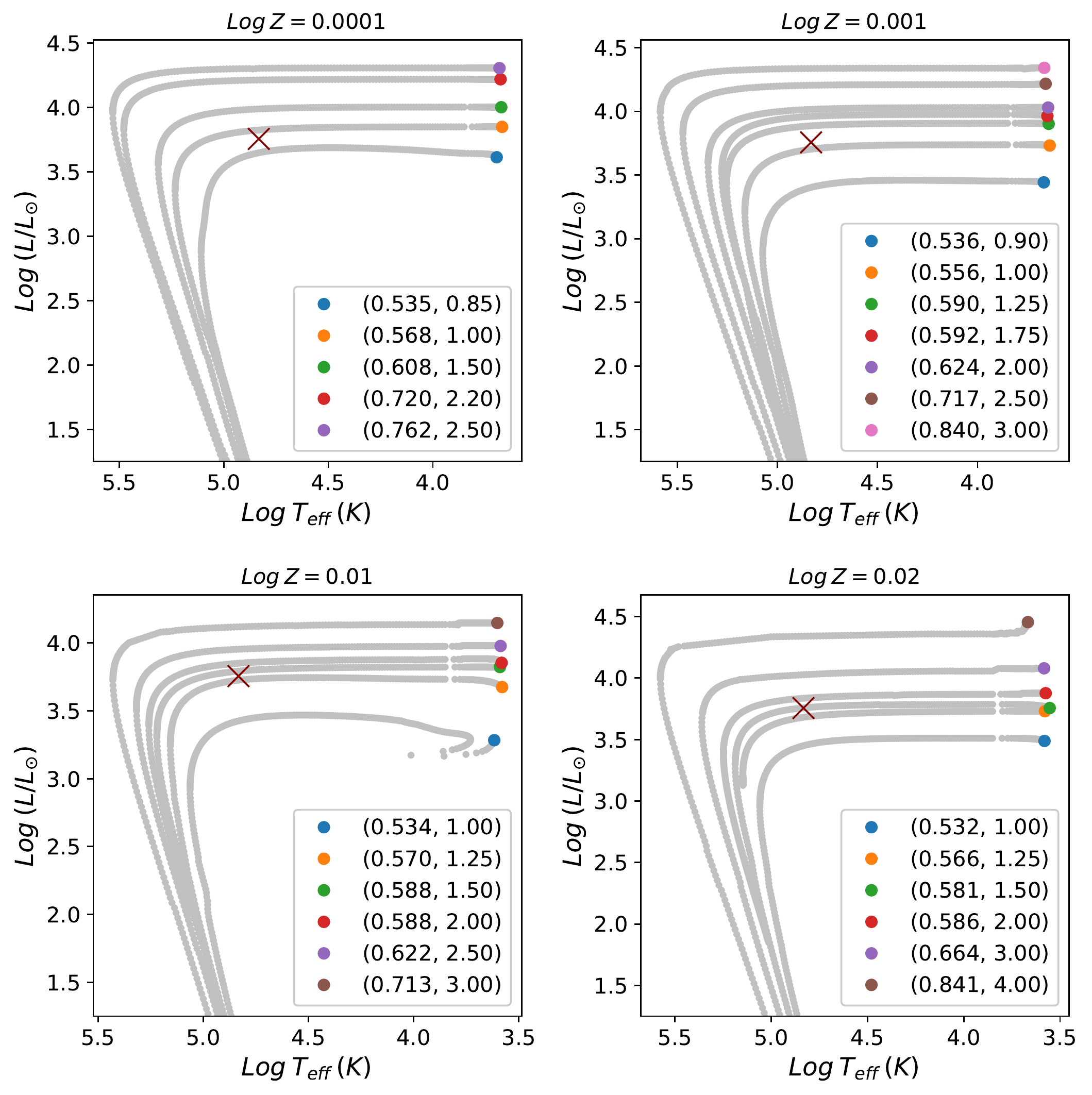}}
 \caption{The evolutionary model tracks (Log $(L/L_{\sun})$ vs. Log $(T_\mathrm{eff})$), calculated by \citet{2016A&A...588A..25M} for different metallicities. Each model track (shown in black) corresponds to $M_\mathrm{rem}$ and $M_\mathrm{pr}$ (see Sec. \ref{sec:bestmodel}). The tracks are labelled as ($M_\mathrm{rem}$, $M_\mathrm{pr}$) with coloured circles at the tip of each track. We place NGC 6572 among the model tracks at the co-ordinates (Log $(L/L_{\sun})$, Log $(T_\mathrm{eff})$) corresponding to our estimated values of $T_\mathrm{eff}$ and $L$ from photoionization modeling. The position of NGC 6572 is marked with red cross in each image box.}
 \label{fig:massfunction}
\end{figure*}

\begin{figure}
\centering
\scalebox{0.55}[0.55]{\includegraphics{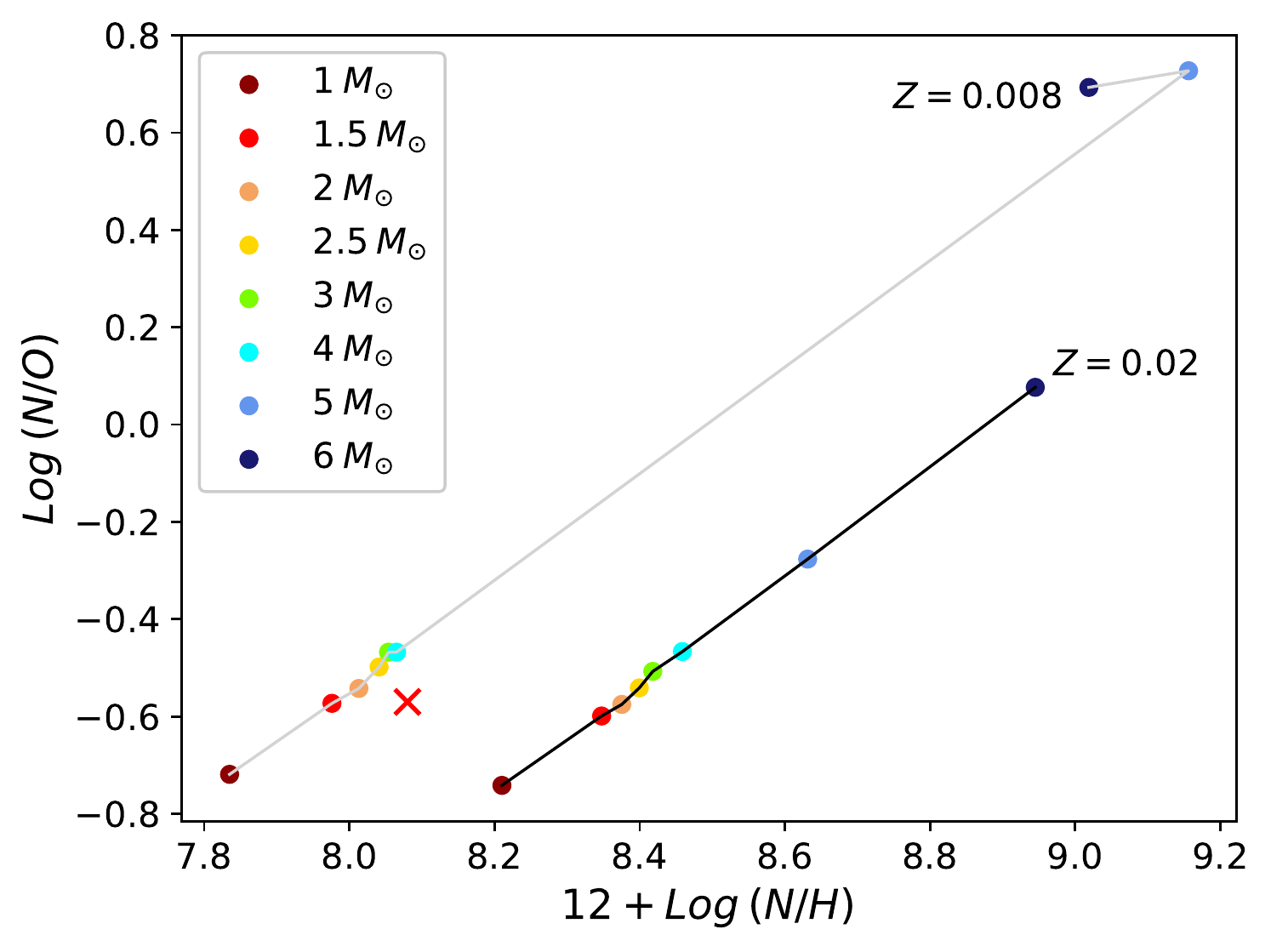}}
 \caption{N/O vs N/H obtained from our photoionization model (red cross) plotted with the model calculated evolutionary yields obtained by \citet{2010MNRAS.403.1413K} for $Z=0.008$ and $Z=0.02$ corresponding to different main-sequence masses (shown in circles for different masses; see figure legend). The comparison gives $M_\mathrm{pr}\sim1.75\:M_{\sun}$ for NGC 6572.}
 \label{fig:yield}
\end{figure}

To estimate the mass of the central star and the progenitor, we use the evolutionary models{\footnote{\url{https://cdsarc.cds.unistra.fr/ftp/J/A+A/588/A25/}}} calculated by \citet{2016A&A...588A..25M} (Fig. \ref{fig:massfunction}). Each model track corresponds to a particular progenitor mass  ($M_\mathrm{pr}$) and calculated with an initial mass equal to the remnant mass ($M_\mathrm{rem}$) at the tip of AGB phase. Using the values of $T_\mathrm{eff}$ and $L$ estimated from our photoionization model, we can estimate $M_\mathrm{rem}\sim0.565$ $M_{\sun}$, from the model tracks. Further, NGC 6572 seem to descend from a low-mass progenitor, with $M_\mathrm{pr}\sim1.25$ $M_{\sun}$. Parallely, we use the N/H and O/H obtained from our photoionization model to estimate $M_\mathrm{pr}$ from the models calculated by \citet{2010MNRAS.403.1413K} (Fig. \ref{fig:yield}). This leads to a higher value of $M_\mathrm{pr}\sim1.75$ $M_{\sun}$ for NGC 6572. Although, it can be noted from Fig. \ref{fig:massfunction} and Fig. \ref{fig:yield} that a central star with $\sim5\%$ higher luminosity and $\sim5\%$ lower N/O would have predicted similar $M_\mathrm{pr}$ from both the methods. Hence, the difference in the estimation of $M_\mathrm{pr}$ from the two methods might be looked as a result of reasonable error in the physical quantities predicted from photoionization modeling.

\subsubsection{Constraining the photoionization model} \label{sec:constrain}

In this work, we mainly focus to constraint the system parameters related to PN, which directly helps to test evolutionary models. We { assume} the previous reporting of the { distances} to NGC 6572 are well-estimated. Most recently, \citet{2021AJ....161..147B} estimated $d=1.77\pm0.15$ kpc for NGC 6572 from accurate GAIA\footnote{\url{https://gea.esac.esa.int/}} measurements. Hence, we narrowed down the parameter range of $d$ in our models to $1.5-2$ kpc. In general, $d$ might be affected due to model degeneracies and often estimated independently (e.g, \citealt{2009A&A...507.1517M}; \citealt{2019MNRAS.482.2354O}; \citealt{10.1093/mnras/staa1518}). Clumpiness of the nebular structure, given by $f$, might induce further uncertainties \citep{2009A&A...507.1517M}, which is often reduced by using $f=1$ in the models. However, the shells of NGC 6572 are evidently clumpy that do not allow us to presume a fixed $f=1$. However, we limited the range of $f$ by running models only with 0.1, 0.5 and 1.0. We find that $f=0.1$ largely underestimates the absolute flux in the model. By a qualitative analysis, we find that $f=0.5$ model reproduces the line flux ratios more consistently than the $f=1.0$ model, although $f=1.0$ reproduces the absolute nebular flux better that the $f=0.5$ model. However, considering the complexity in the morphology of the shells of NGC 6572, it is likely that the nebular shell might have additional components of similar ionization structure, which are not considered in this model. Those missing components might contribute to the total flux. Hence, our final model is attributed with $f=0.5$.
In this work, with a pre-estimated narrow parameter range for $d$, and with the inter-dependence of $d$, shell radii and a well-justified value of $f$, we put some a-priori constraint on the model. Finally, the model is validated since the observables are well-reproduced. Furthermore, we may claim the constrain is even better due to detailed treatment of nebular morphology in the photoionization modelling.

\subsubsection{Evaluation of the fitting of fluxes}  \label{sec:evaluation}

Our photoionization model well-reproduce the line ratios governing $T_\mathrm{e}$, such as [O~{\sc III}] 4363/5007, [N~{\sc II}] 5755/6584, and [S~{\sc III}] 6312/9069, except for [S~{\sc III}] 6312/9069 in the modelled NS spectrum, where the difference from the observation is about $30\%$. The line ratios characterizing $N_\mathrm{e}$, [S~{\sc iii}] 6716/6731 and [Ar~{\sc iv}] 4740/4711 are close to their observed values in the photoionization model. The modelled [Cl~{\sc iii}] 5538/5518 have around 35$\%$ deviation form the observed ratio. 

Our model cannot properly generate the He~{\sc ii} 4686 {\AA} flux. However, we find that the central star of NGC 6572 is reportedly a weak emission line star (wels) \citep{2020MNRAS.493..730M} that may have a stellar He~{\sc ii}. \citet{1988A&A...207L...5M} reported the first evidence of nebular He~{\sc ii} generation in high-resolution spectra of NGC 6572.
We note that $I$(He~{\sc ii} 4686/$H\beta)$ for NGC 6572 from low-resolution spectra (including ours) $\sim0.008-0.01$, while the high-resolution observations of the PN give nebular $I$(He~{\sc ii} 4686/$H\beta)<0.004$. Hence, our observed He~{\sc ii} 4686 {\AA} flux could be the sum of the stellar and nebular component. 

The model closely reproduces the $H\beta$ flux obtained in the EW spectra. The modelled $H\beta$ flux is about $30\%$ lower than the observed value. This seems justifiable, since the photoionization model uses a simplified bipolar shell, whereas NGC 6572 depicts complex stuctures with multiple lobes that may account for the deficit of $H\beta$ flux in the model. 

The abundances estimated from our photoionization model deviate up to $\sim25\%$ from the abundances calcuated using ICFs, since it was necessary to change the initial abundances (equal to the total abundances obtained using ICFs) of the elements during the modelling to match the corresponding line fluxes. It is to be noted that a photoionization model simultaneously predict all the ionization states of an element, hence, predicts the total abundances self-consistently. Here, the abundances may be considered to be improved through photoionization modelling. 

\subsubsection{Fitting of resolved emission line profiles}

\begin{figure}
\centering
\scalebox{0.5}[0.5]{\includegraphics{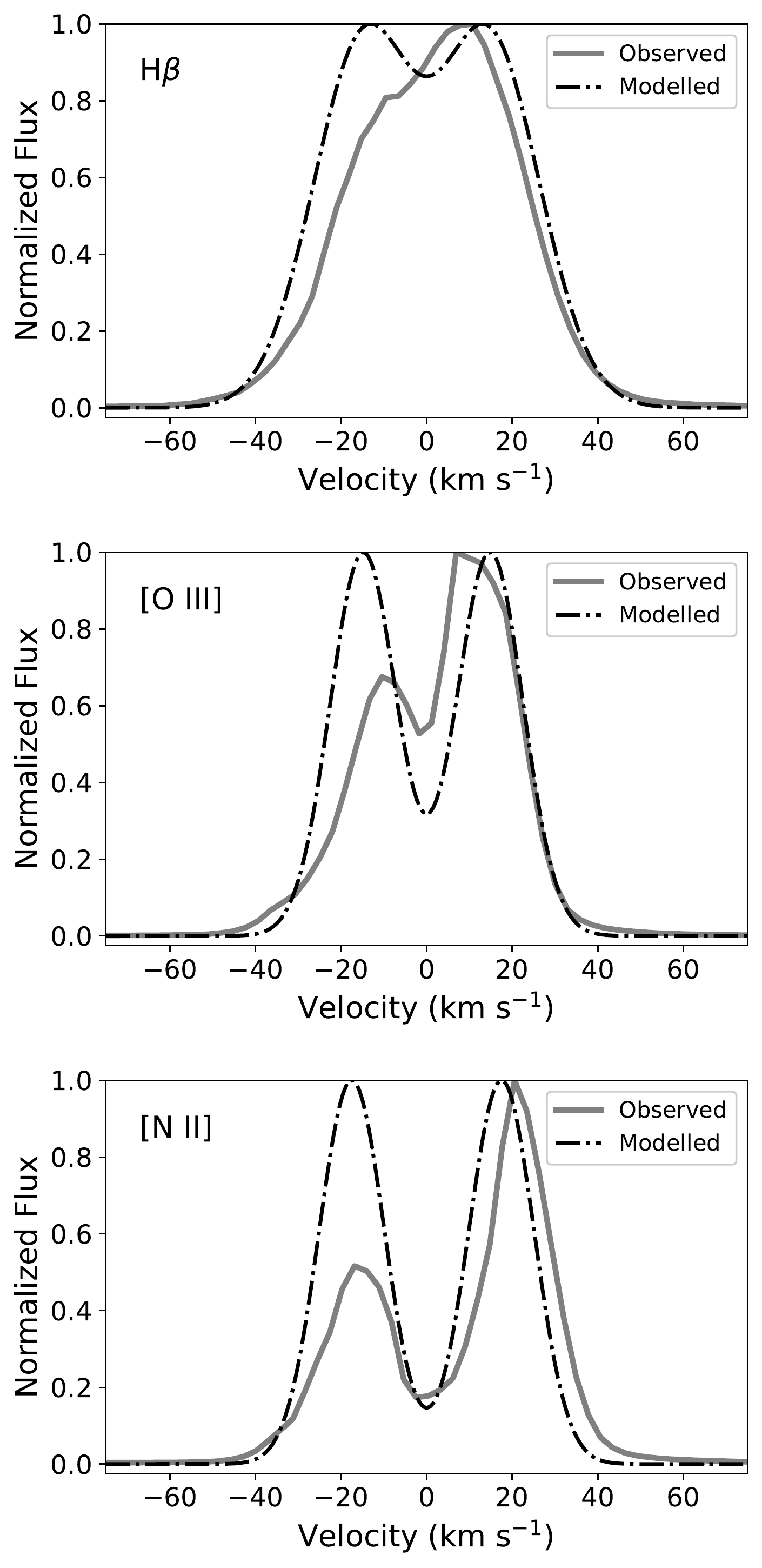}}
 \caption{The \textsc{pycloudy} fitting of H$\beta$ (top), [O~{\sc iii}] (middle), and [N~{\sc ii}] (bottom) velocity profiles obtained from WHT spectra. \label{fig:WHTprofiles}}
\end{figure}

We include a velocity field parameter (expansion velocity of the nebula) in our photoionization model to obtain the kinematic signatures in the modelled emission line profiles. We attempt to fit the resolved high-resolution spectral profiles of H$\beta$, [O~{\sc iii}], and [N~{\sc ii}] observed in the WHT spectra with the model generated profiles (Fig. \ref{fig:WHTprofiles}). We observe that the observed profiles have structural differences, which is due to the difference in the distribution of the ionization states of the elements. We notice that our modelled profiles show the similar characteristic differences in the H$\beta$, [O~{\sc iii}], and [N~{\sc ii}] profiles. The expansion velocity of the nebula has been adjusted to match the width of the profiles.  

\section{Summary and conclusion}

In this paper, we study the planetary nebula NGC 6572, a PN with multipolar morphology and moderate ionization. A 3D morpho-kinematic model is reconstructed with the reference of \textit{HST} image and PV diagrams. The modelled image and PV diagrams are satisfactorily reproduced. We obtain the 3D morphology comprising structural components: bipolar shells and toroidal waist. We obtain the relative orientation and densities, and velocity fields of the components. We estimate the inclination of the nebula along the line-of-sight. 

Using \textsc{pyneb}, We derive the plasma parameters ($T_\mathrm{e}$ and $N_\mathrm{e}$) from characteristic spectral emission line flux ratios. The ionic and total elemental abundances are also derived directly from the emission lines. To study the ionization structure, we compute a pseudo-3D photoionization model of the PN using \textsc{cloudy} and the \textsc{pycloudy} library. The input nebular shell configuration for the photoionization model is guided by the shell structure obtained from the 3D model. We fit the absolute flux and flux ratios in the observed spectra with the modelled absolute flux and line ratios. We self-consistently estimate the physical  parameters: central star effective temperature, luminosity and gravity; nebular shell geometry, density, and elemental abundances. 

We estimate the progenitor mass and remnant mass using stellar evolutionary trajectories. The model satisfactorily reproduces the fluxes and the characteristic line ratios of the observed spectra. We also compare the resolved emission line profiles of H$\beta$, [O~{\sc iii}], and [N~{\sc ii}] with modelled line profiles obtained from the photoionization model. We observe that the kinematic signature in the line profiles are well-reproduced. The morpho-kinematic and photoionization models are well-connected in morphological aspect, which is a major improvement in the methodology than the previous modelling approaches for this PN. Further improvements could be done in the 3D morpho-kinematic modelling of the filamentary micro-structures in the shells. However, this approach will require sufficiently resolved PV diagrams of the lobes.
        
\section*{Acknowledgements}
We thank Dr. Jesus A. Toala for valuable comments and suggestions that helped to improve this paper.
We acknowledge S. N. Bose National Centre for Basic Sciences under Department of Science and Technology (DST), Govt. of India, for providing necessary support to conduct research work. We are thankful to the HCT Time Allocation Committee (HTAC) for allocating nights for observation, and the supporting staff of the observatory. This paper uses data based on observations made with the NASA/ESA Hubble Space Telescope, and obtained from the Hubble Legacy Archive, which is a collaboration between the Space Telescope Science Institute (STScI/NASA), the Space Telescope European Coordinating Facility (ST-ECF/ESA) and the Canadian Astronomy Data Centre (CADC/NRC/CSA). This paper uses archival data from the 4.2 m WHT telescope. We thank the people maintaining these data bases. This work made use of Astropy:\footnote{http://www.astropy.org} a community-developed core Python package and an ecosystem of tools and resources for astronomy \citep{2022ApJ...935..167A}.

\section*{Data Availability}
The optical spectra used in this paper will be shared on reasonable request to the corresponding author. The \textit{HST} data are available at the Hubble Legacy Archive (HLA; https://hla.stsci.edu/). 

\bibliographystyle{mnras}
\bibliography{References}


\bsp	
\label{lastpage}
\end{document}